\documentclass[twocolumn]{aastex631}

\usepackage{amsmath}
\shorttitle{Chemical Signatures of an Embedded Planet in the HD~169142 Disk}
\shortauthors{Law et al.}
\graphicspath{{./}{figures/}}

\begin{document}

\title{SO and SiS Emission Tracing an Embedded Planet and Compact $^{12}$CO and $^{13}$CO Counterparts in the HD~169142 Disk}

\author[0000-0003-1413-1776]{Charles J. Law}
\affiliation{Center for Astrophysics \textbar\, Harvard \& Smithsonian, 60 Garden St., Cambridge, MA 02138, USA}

\author[0000-0003-2014-2121]{Alice S. Booth}
\affiliation{Leiden Observatory, Leiden University, 2300 RA Leiden, the Netherlands}

\author[0000-0001-8798-1347]{Karin I. \"Oberg}
\affiliation{Center for Astrophysics \textbar\, Harvard \& Smithsonian, 60 Garden St., Cambridge, MA 02138, USA}



\begin{abstract}
Planets form in dusty, gas-rich disks around young stars, while at the same time, the planet formation process alters the physical and chemical structure of the disk itself. Embedded planets will locally heat the disk and sublimate volatile-rich ices, or in extreme cases, result in shocks that sputter heavy atoms such as Si from dust grains. This should cause chemical asymmetries detectable in molecular gas observations. Using high-angular-resolution ALMA archival data of the HD~169142 disk, we identify compact SO J=8$_8$--7$_7$ and SiS J=19--18 emission coincident with the position of a ${\sim}$2~M$_{\rm{Jup}}$~planet seen as a localized, Keplerian NIR feature within a gas-depleted, annular dust gap at ${\approx}$38~au. The SiS emission is located along an azimuthal arc and has a similar morphology as a known $^{12}$CO~kinematic excess. This is the first tentative detection of SiS emission in a protoplanetary disk and suggests that the planet is driving sufficiently strong shocks to produce gas-phase SiS. We also report the discovery of compact $^{12}$CO and $^{13}$CO~J=3--2 emission coincident with the planet location. Taken together, a planet-driven outflow provides the best explanation for the properties of the observed chemical asymmetries. We also resolve a bright, azimuthally-asymmetric SO ring at ${\approx}$24~au. While most of this SO emission originates from ice sublimation, its asymmetric distribution implies azimuthal temperature variations driven by a misaligned inner disk or planet-disk interactions. Overall, the HD~169142 disk shows several distinct chemical signatures related to giant planet formation and presents a powerful template for future searches of planet-related chemical asymmetries in protoplanetary disks.
\end{abstract}
\section{Introduction} \label{sec:intro}
Planets form and inherit their compositions in dusty, gas-rich disks around young stars, while the planet formation process is expected to simultaneously alter the physical and chemical structure of the disk. Observations of protoplanetary disks using the Atacama Large Millimeter/submillimeter Array (ALMA) have revealed the presence of ubiquitous rings and gaps in the millimeter dust distribution \citep[][]{ALMA15,Andrews18, Huang18,Cieza21} and in molecular line emission \citep[][]{Bergner19, vanTerwisga19, Facchini21PDS, Oberg21_MAPSI, LawMAPSIII}. In some cases, the millimeter continuum and molecular line emission appear azimuthally asymmetric \citep{Tang12, vanderMarel21, vanderMarel21_dust, Booth21_IRS, Booth23}. While such features broadly suggest the presence of embedded planets, it remains difficult to unambiguously connect individual substructures seen in dust and line emission with the location and properties of nascent planets.

Nonetheless, a variety of methods for identifying and characterizing embedded planets in disks have been developed. Planet formation is thought to occur within dust gaps and the properties of the observed dust distributions have been used to infer planet characteristics \citep[e.g.,][]{Kanagawa15, Zhang18}, while in a few systems, circumplanetary disks (CPDs) are detected via their excess millimeter continuum emission \citep{Keppler18, Benisty21, Wu22}. However, such detections remain relatively scare \citep{Isella14, Pineda19, Andrews21}, which may be due to mm dust depletion and high gas-to-dust ratios (${>}$1000) in CPDs \citep{Karlin23}. Thus, the identification of planetary signals in molecular gas provides a promising complementary approach. While planetary signatures are now often indirectly seen via kinematic perturbations in the rotation velocity profiles of bright lines \citep[e.g.,][]{Teague18_HD163296, Pinte20}, the first CPD candidate in molecular gas was discovered in $^{13}$CO gas in the AS~209 disk \citep{Bae21}. A similar planetary counterpart was recently identified in $^{12}$CO gas in the Elias 2-24 disk \citep{Pinte23}. 

Embedded planets are also expected to directly alter the chemical structure of disks. In particular, forming-planets should locally heat the disk \citep[e.g.,][]{Szulagyi17, Szulagyi18}, which will sublimate volatile-rich ice, or in more extreme cases, drive shocks that sputter heavy atoms such as Si from dust grains. This will result in chemical asymmetries that can be detected in sensitive line emission observations \citep{Cleeves15, Rab19}. Sufficiently deep observations of molecular gas with ALMA have only recently become available to enable searches for such chemical signatures. For instance, asymmetric SO line emission was found to trace an embedded planet in the HD~100546 disk \citep{Booth23} and \citet{Alarcon22} identified a signature in CI gas potentially attributable to either the inflow or outflow associated with a protoplanet in the HD~163296 disk.

In this Letter, we present high-angular-resolution ALMA archival observations of the HD 169142 disk. We identify several chemical signatures related to ongoing planet formation, including spatially-localized SO and SiS emission and the discovery of compact $^{12}$CO and $^{13}$CO emission counterparts, which are all approximately coincident with an embedded giant planet. In addition, spatially-resolved SO emission has only been detected in a handful of Class II disks to date \citep{Pacheco16, Booth18, Booth21_IRS, Booth23, Huang23} and we report the first tentative detection of SiS in a protoplanetary disk. In Section \ref{sec:HD169142_disk}, we describe the HD~169142 disk and present the ALMA observations in Section \ref{sec:observations_overview}. We present our results in Section \ref{sec:results} and discuss the chemical origins of the observed emission and possible connections with nascent planets within the HD~169142 disk in Section \ref{sec:discussion}. We summarize our conclusions in Section \ref{sec:conlcusions}.

\section{The HD~169142 Disk and Evidence for Embedded Planets}  \label{sec:HD169142_disk}

HD~169142 is a Herbig Ae star \citep{Blondel06} located at a distance of $d=115$~pc \citep{Gaia21, Bailer_Jones21} and has a dynamically-estimated stellar mass of $M_*=1.4~\rm{M}_*$ \citep{Yu21}, luminosity of $L_*=10~\rm{L}_{\odot}$ \citep{Fedele17}, and age of ${\sim}$10~Myr \citep{Pohl17}. The disk surrounding HD~169142 has a nearly face-on orientation with an inclination of $i=13^{\circ}$ and position angle of PA$=5^{\circ}$ \citep{Raman06, Panic08}.

The HD~169142 disk consists of several ring-like structures observed across many wavelengths, including in NIR/scattered light \citep[e.g.,][]{Quanz13, Reggiani14, Pohl17, Ligi18, Bertrang18, Gratton19}, thermal mid-infrared \citep{Honda12}, (sub)-mm \citep[e.g.,][]{Fedele17, Macias19, Perez19}, and cm \citep{Osorio14}. Many of these structures have been posited to originate from planet-disk interactions. While varying numbers and masses of planets have been suggested, most studies require at least one ${\gtrsim}$M$_{\rm{Jup}}$ planet to explain the observed disk substructures. For a detailed summary of the proposed locations and types of embedded planets, please see \citet{Yu21} and \citet{Garg22}, and references therein. Here, we restrict our focus to the annular gap located at an approximate radius of 41~au (``D41") in-between two major rings at ${\approx}$25~au (``B25") and ${\approx}$60~au (``B60"). The D41 gap is observed in both mm continuum and NIR/scattered light.

\setlength{\tabcolsep}{3.5pt}
\begin{deluxetable*}{lccccccccccccccc}[!]
\tabletypesize{\footnotesize}
\tablecaption{Image Cube and Line Properties\label{tab:image_info}}
\tablehead{
\colhead{Transition} & \colhead{Freq.} & \colhead{Beam} & \colhead{JvM $\epsilon$\tablenotemark{a}} & \colhead{\texttt{robust}} & \colhead{Chan. $\delta$v} & \colhead{RMS} & \colhead{Project} & \colhead{E$_{\rm{u}}$} & \colhead{A$_{\rm{ul}}$} & \colhead{g$_{\rm{u}}$} & \colhead{Int. Flux\tablenotemark{b}}  \\ 
\colhead{} & \colhead{(GHz)} & \colhead{ ($^{\prime \prime} \times ^{\prime \prime}$, $\deg$)} & & & \colhead{(km~s$^{-1}$)} & \colhead{(mJy~beam$^{-1}$)} & \colhead{Used} & \colhead{(K)} & \colhead{($\log_{10}$ s$^{-1}$)} & & \colhead{(mJy~km~s$^{-1}$}) }
\startdata
0.9~mm cont.        & 331.000000 & 0.05 $\times$ 0.03, 79.6 & 0.42  & $-$0.5 & \ldots & 0.1 & P2012$+$P2015 & \ldots & \ldots & \ldots &  420 $\pm$ 10 \\
$^{12}$CO J=3--2        & 345.795990 & 0.11~$\times~$0.09, 74.2 & 0.64 & 0.5 & 0.12 & 3.9 & P2012$+$P2015 & 33 & $-$5.603 & 7 & 27958 $\pm$ 228 \\
$^{13}$CO J=3--2 & 330.587965 & 0.14~$\times$~0.11, 75.1 & 0.67 & 0.5 & 0.12 & 3.1 & P2012$+$P2015 & 32 & $-$5.960 & 14 & 10664 $\pm$ 205  \\
SO J=8$_{8}$--7$_{7}$ & 344.310612 & 0.19~$\times$~0.14, 84.4 & \ldots & 2.0 & 0.43 & 1.5 & P2012 & 88 & $-$3.285 &17 & 120 $\pm$ 16 \\
SO$_2$ J=4$_{3,1}$--3$_{2,2}$ & 332.505242 & 0.20~$\times$~0.15, 88.2 & \ldots & 2.0 & 0.44 & 1.4 & P2012 & 31 &  $-$3.483 & 9 & ${<}$32 \\
SO$_2$ J=11$_{6,6}$--12$_{5,7}$ & 331.580244 & 0.20~$\times$~0.15, 88.2 & \ldots & 2.0 & 0.44 & 1.5 & P2012 & 149 & $-$4.362 & 23 & ${<}$43 \\ 
SiS J=19--18 & 344.779481 & 0.19~$\times$~0.14, 84.3 & \ldots & 2.0 & 0.43 & 1.6 & P2012 & 166 & $-$3.155 & 39 & 62 $\pm$ 23  \\
\enddata
\tablecomments{The spectroscopic constants for all lines are taken from the CDMS database \citep{Muller01, Muller05, Endres16}.}
\tablenotetext{a}{The ratio of the CLEAN beam and dirty beam effective area used to scale image residuals to account for the effects of non-Gaussian beams. See Section \ref{sec:selfcal_imaging} and \citet{JvM95, Czekala21} for further details.}
\tablenotetext{b}{Uncertainties are derived via bootstrapping and do not include the systematic calibration flux uncertainty (${\sim}$10\%). 3$\sigma$ upper limits are reported for nondetections. The continuum flux has units of mJy.}
\end{deluxetable*} \vspace{-24pt} \setlength{\tabcolsep}{4pt}

Multiple lines of evidence point to a planetary origin of the D41 gap, including hydrodynamic simulations \citep{Bertrang18, Toci20}; a decreased gas surface density \citep{Fedele17, Garg22} and localized kinematic excess in $^{12}$CO within this gap \citep{Garg22}. The most direct evidence of an embedded planet is in the form of NIR observations, which revealed a localized emission feature within this annular gap that is in Keplerian rotation around the central star as well as the presence of a spiral-like structure consistent with a planet-driven wake \citep{Gratton19, Hammond23}. This feature was suggested to correspond to a ${\sim}$2~M$_{\rm{Jup}}$ still-accreting planet at a radius of ${\approx}$38~au \citep{Gratton19}. Taken together, this planet may be responsible for carving this gas-depleted, annular gap; exciting a spiral-wake in the NIR; and triggering the observed $^{12}$CO kinematic excess. Overall, this makes the HD~169142 disk an ideal source to search for chemical signatures associated with an embedded giant planet.

\section{Observations}
\label{sec:observations_overview}

\subsection{Archival Data and Observational Details}
\label{sec:archival_data}

We made use of ALMA archival projects 2012.1.00799.S (PI: M. Honda) and 2015.1.00806.S (PI: J. Pineda), which we hereafter refer to as P2012 and P2015, respectively. P2012 comprises three execution blocks with baselines ranging from 15-1574~m, and P2015 has a single execution block with baselines of 19-7716~m. Table \ref{tab:full_obs_program_details} provides detailed information about each project.

Both projects had two narrow spectral windows (244.1~kHz; ${\approx}$0.12~km~s$^{-1}$) centered on the $^{12}$CO J=3--2 and $^{13}$CO J=3--2 lines. In addition, P2012 had two spectral windows covering the frequency ranges of 331.055-332.93~GHz and 343.055-344.929~GHz but at a coarse velocity resolution (976.6~kHz; ${\approx}$0.45~km~s$^{-1}$). P2015 instead included one spectral window from 342.555-344.429~GHz but at an even coarser velocity resolution (1,128.9~kHz; ${\approx}$0.85~km~s$^{-1}$) and a dedicated continuum window. Within these spectral windows, we imaged the following lines: $^{12}$CO J=3--2, $^{13}$CO J=3--2, SO J=8$_{8}$--7$_{7}$, SO$_2$ J=4$_{3,1}$--3$_{2,2}$, SO$_2$ J=11$_{6,6}$--12$_{5,7}$, and SiS J=19--18. Table \ref{tab:image_info} lists details for each image cube. 

The CO and SO lines are included in spectral set-ups in both P2012 and P2015. For the CO lines, we use both projects since each had dedicated spectral windows with high velocity resolution. For SO, we initially imaged the line using both the P2012-only and P2012+P2015 data and confirmed it is robustly detected in both combinations. However, due to the considerably coarser channel spacing of the P2015 data (${\approx}$0.85 km~s$^{-1}$), we performed all subsequent analysis of SO using only the P2012 data. The SiS and SO$_2$ lines are only covered in P2012. We generated the 0.9~mm continuum image using data from both projects and used the full bandwidth of the observations after flagging the channels containing line emission. Table \ref{tab:image_info} lists which project was used to generate each image cube.

Transitions of additional S- and Si-bearing molecules of interest (e.g., SiO, CCS, OCS, NS) were not included in the spectral set-up of either project.

\subsection{Self Calibration and Imaging}
\label{sec:selfcal_imaging}

Each archival project was initially calibrated by ALMA staff using the ALMA calibration pipeline and the required version of CASA \citep{McMullin_etal_2007}, before switching to CASA \texttt{v5.4.0} for self calibration. Self calibration was attempted on the shorter baseline data from P2012 alone but we were unable to derive solutions that improved image quality. Data from both projects were combined and one round of phase self-calibration was applied, following the same procedure described in \citet{Oberg21_MAPSI}. We subtracted the continuum using the \texttt{uvcontsub} task with a first-order polynomial. 

\begin{figure*}
\centering
\includegraphics[width=0.925\linewidth]{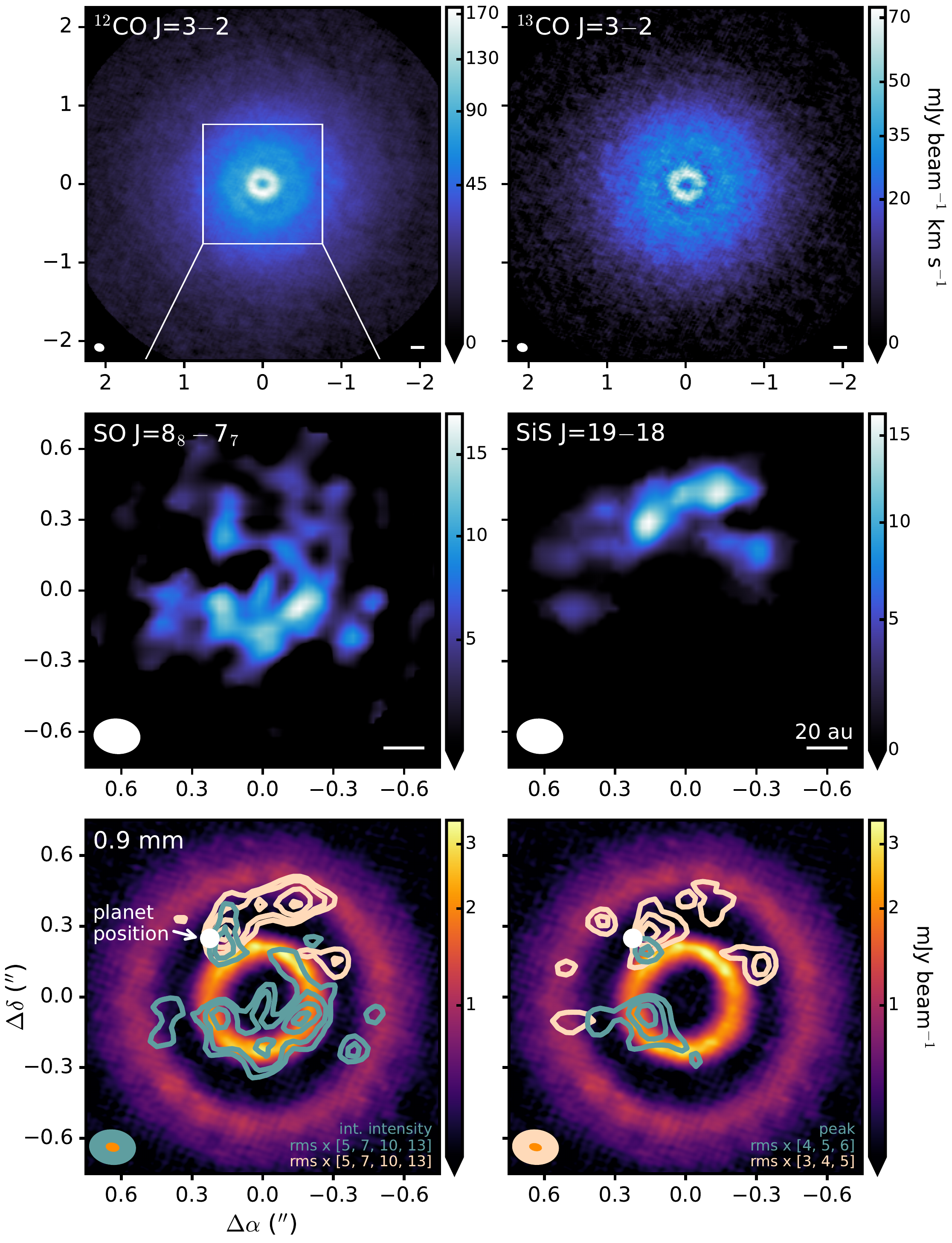}
\caption{Top two rows: zeroth moment maps of $^{12}$CO J=3--2, $^{13}$CO J=3--2, SO J=8$_8$--7$_7$, SiS J=19--18 for the HD~169142 disk. The synthesized beam and a scale bar indicating 20~au is shown in the lower left and right corner, respectively, of each panel. Bottom row: composite image of the 0.9~mm continuum emission (colorscale) with contours showing the integrated (left) and peak intensity (right) of SO and SiS. The RMS for the integrated intensities represents the zeroth moment map uncertainty and for the peak intensities, the RMS is the channel map noise. The planet location from \citet{Gratton19, Hammond23} is marked by a white circle. The synthesized beams for the continuum (orange), SO (teal), and SiS (peach) are shown in the lower left corners. The disk rotation is clockwise.} 
\label{fig:figure1}
\end{figure*}

We then switched to CASA \texttt{v6.3.0} for all imaging. We used the \texttt{tclean} task to produce images of all transitions with Briggs weighting and Keplerian masks generated with the \texttt{keplerian\_mask} \citep{rich_teague_2020_4321137} code. Each mask was based on the stellar and disk parameters of HD~169142 and was visually inspected to ensure that it contained all emission present in the channel maps. Due to the non-Keplerian nature of the SiS emission, no masking was used to generate the SiS images. Briggs \texttt{robust} parameters were chosen manually to prioritize high signal-to-noise ratio (SNR) line detections. Channel spacings ranged from 0.12-0.85~km~s$^{-1}$ depending on the line. All images were made using the ‘multi-scale’ deconvolver with pixel scales of [0,5,15,25] and were CLEANed down to a 4$\sigma$ level, where $\sigma$ was the RMS measured in a line-free channel of the dirty image. Table \ref{tab:image_info} summarizes all image properties.

For the $^{12}$CO, $^{13}$CO, and continuum images, we applied the `JvM' correction proposed in \citet{Jorsater95} and described in more detail in \citet{Czekala21}. This correction scales the image residuals by a factor $\epsilon$, equal to the ratio of the effective areas of the CLEAN beam and dirty beam, to be in units consistent with the CLEAN model. Table \ref{tab:image_info} lists all $\epsilon$ values. 

\subsection{Moment Maps, Radial Profiles, and Integrated Fluxes}
We generated velocity-integrated intensity, or ``zeroth moment," maps of line emission from the image cubes using \texttt{bettermoments} \citep{Teague18_bettermoments} and closely followed the procedures outlined in \citet{LawMAPSIII}. No flux threshold for pixel inclusion, i.e., sigma flipping, was used to ensure accurate flux recovery. All maps were generated using the same Keplerian masks employed during CLEANing, except in the case of SiS, where we used hand-drawn masks to better capture the non-Keplerian emission (see Figure \ref{fig:figure_appendix_sis} in the Appendix). \texttt{bettermoments} also provides a map of the statistical uncertainty that takes into account the non-uniform RMS across the zeroth moment map, which is described in detail in \citet{Teague19RNAAS}. We take the median RMS from these uncertainty maps as the zeroth moment map uncertainty. We also generated peak intensity maps using the ‘quadratic’ method of \texttt{bettermoments}.

We computed radial line intensity profiles using the \texttt{radial\_profile} function in the \texttt{GoFish} python package \citep{Teague19JOSS} to deproject the zeroth moment maps. All radial profiles assume a flat emitting surface. We computed the integrated fluxes for all lines with the same masks used to generate the zeroth moment maps. We estimated uncertainties as the standard deviation of the integrated fluxes within 500 randomly-generated masks at the same spatial position but spanning only line-free channels. We report $3\sigma$ upper limits for the two undetected SO$_2$ lines. Table \ref{tab:image_info} lists all integrated fluxes.

\section{Results} \label{sec:results}

\subsection{Spatial Distribution of Emission} \label{sec:line_emission}

\subsubsection{CO Emission Morpohology} \label{sec:CO_morpohology}

Figure \ref{fig:figure1} shows zeroth moment maps of the detected lines. The $^{12}$CO and $^{13}$CO J=3--2 lines are the most spatially extended and display two emission rings, closely resembling the J=2--1 transitions presented in \citet{Yu21} and \citet{Garg22} and the $^{13}$CO J=6--5 line from \citet{Leemker22}. In this Letter, we do not analyze the large-scale CO emission in detail and instead refer the interested reader to the above references.

\subsubsection{SO Emission Morpohology} \label{sec:SO_morpohology}

Figure \ref{fig:figure2} shows an azimuthally-averaged SO radial profile. The bulk of the SO emission is located in an azimuthally-asymmetric ring at a radius of ${\approx}$24~au, which closely traces the inner mm dust ring (Figure \ref{fig:figure1}). This SO ring exhibits a clear asymmetry with the southern half of the disk showing bright SO emission, while the northern half has only faint emission. Given the current data quality, it is difficult to precisely determine the degree of azimuthal asymmetry but the peak brightness likely varies by at least a factor of several. A fainter outer ring at ${\approx}$50~au is also detected, which is clearly seen in the radial profile (Figure \ref{fig:figure2}) and is visible as diffuse emission in the zeroth moment map (Figure \ref{fig:figure1}). Due its low SNR, we cannot assess if this outer SO ring is also asymmetric. There also appears to be a central dip in SO intensity at ${\approx}$11~au, but the true depth of this feature is difficult to quantify.

In addition to this ring-like morphology, SO shows a point-source-like emission feature at a radius of 32~au and PA$\approx$35$^{\circ}$. This feature is not spatially-resolved and its compact nature is most clearly visible in the peak intensity map (Figure \ref{fig:figure1}) and in the channel maps (Figure \ref{fig:figure_appendix_so} in the Appendix). We also confirm that this feature is robustly detected in images with a range of Briggs \texttt{robust} parameters (see Appendix \ref{sec:appendix_image_tests}). We often colloquially refer to such localized features as emission ``blobs" and designate them by cardinal directions, i.e., ``SO NE blob". In Figure \ref{fig:figure2}, we also extracted a radial profile along a narrow azimuthal wedge containing this feature. The SO NE blob is not part of either SO ring due its distinct radius, and the PA of this feature corresponds to the azimuthal region of the inner SO ring that shows little-to-no emission.

\subsubsection{SiS Detection and Emission Morpohology} \label{sec:SiS_morpohology}

We report the first tentative detection of SiS in a protoplanetary disk. SiS emission is detected with a peak intensity of 7.9~mJy~beam$^{-1}$, which corresponds to a peak SNR of 5$\sigma$ (Figure \ref{fig:figure1}). We also detected peak SiS emission of at least 4$\sigma$ in three velocity channels, two of which are consecutive (see Figure \ref{fig:figure_appendix_sis}). Despite passing traditional thresholds for line detections \citep[e.g.,][]{Bergner19}, we conservatively refer to this as a tentative detection since only one SiS line is covered in these archival data. 

\begin{figure*}[]
\centering
\includegraphics[width=\linewidth]{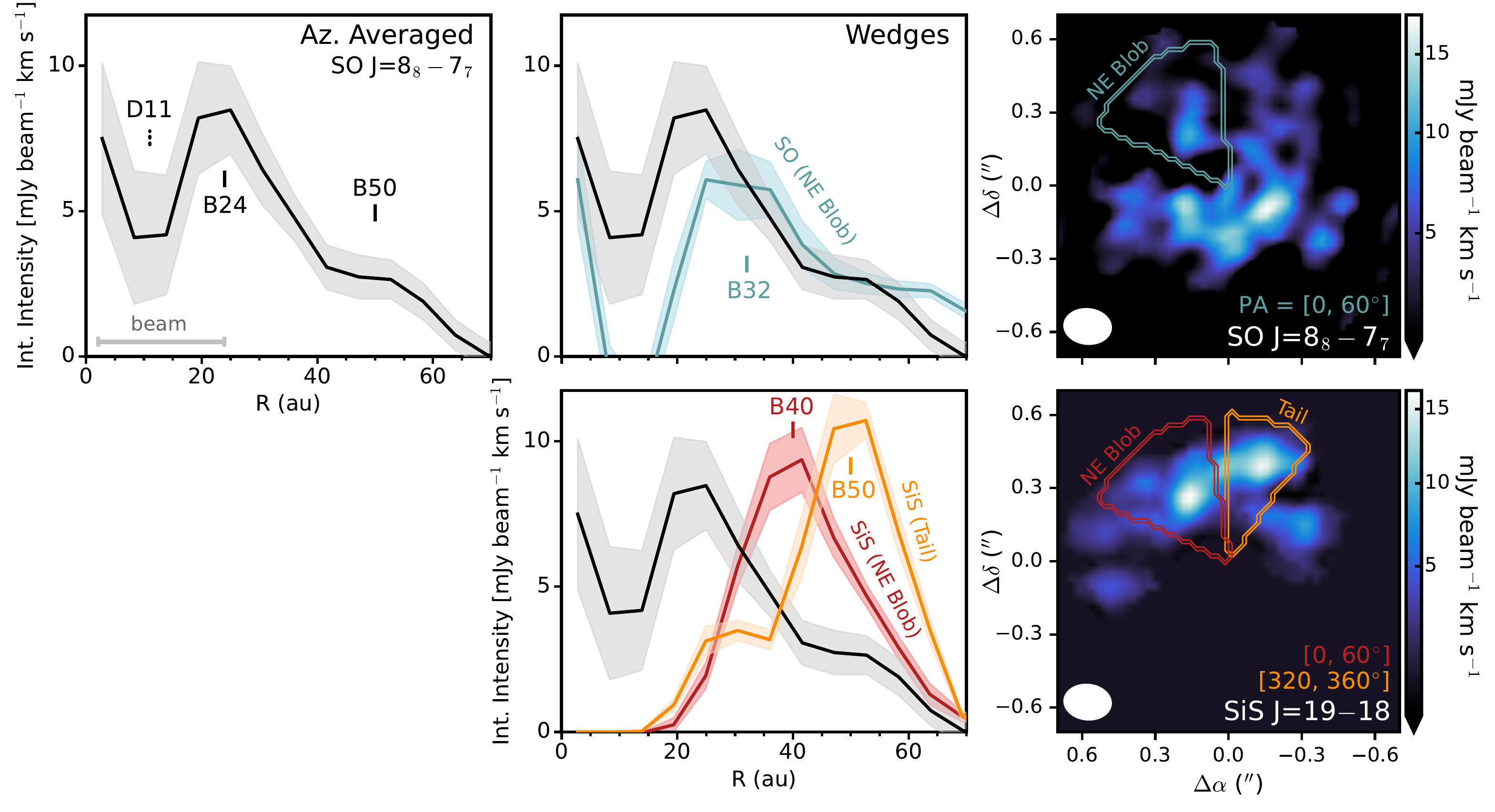}
\caption{Deprojected radial intensity profiles of SO and SiS for the HD~169142 disk (left, middle columns). Shaded regions show the 1$\sigma$ uncertainty. Solid lines mark rings and dotted lines denote gaps. The major axis of the synthesized beam is shown in the lower left corner. SO and SiS emission profiles are extracted from narrow wedges over the labeled PA range, measured east of north in the sky frame. These wedges are illustrated in the associated zeroth moment maps (right column).}
\label{fig:figure2}
\end{figure*}

The SiS emission is spatially-compact, azimuthally-asymmetric, and non-Keplerian. As shown in Figure \ref{fig:figure2}, the SiS emission can be broadly categorized into two distinct components: (1) a localized emission feature (``SiS NE blob") at a similar, but not identical, position (r$\approx$40~au; PA=30$^{\circ}$) as that of the compact SO feature; (2) an extended arc or ``tail" connected which extends to r$\approx$50~au. These features can also clearly be seen in the channel maps in Figure \ref{fig:figure_appendix_sis} in the Appendix. Similar to the SO feature, the SiS NE blob is spatially-unresolved but due to its extended nature, the SiS tail is marginally resolved.

\subsection{Kinematics of SO and SiS} \label{sec:kinematics}

\begin{figure*}
\centering
\includegraphics[width=0.75\linewidth]{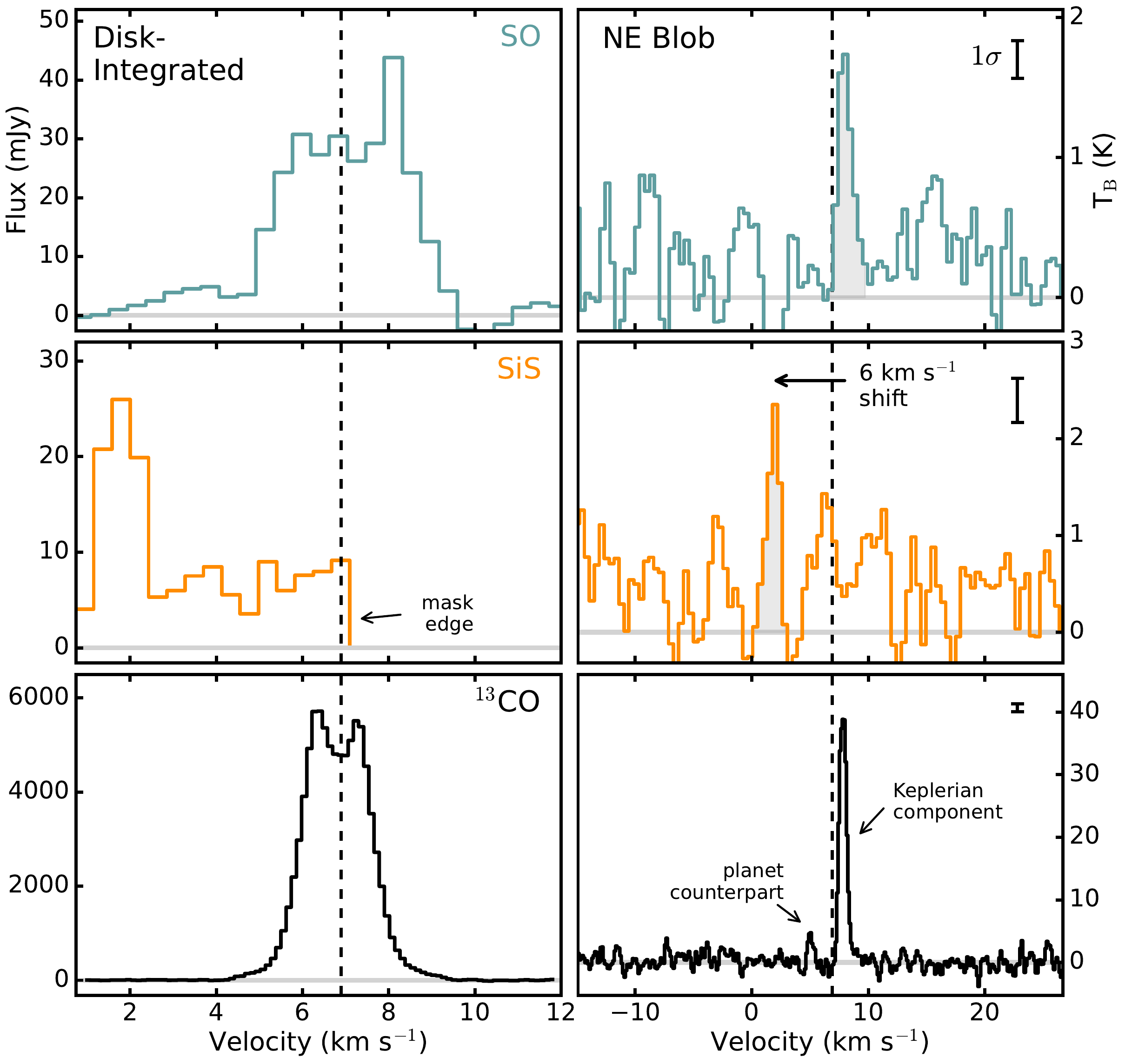}
\caption{Disk-integrated spectra (left) and at the position of the NE Blob extracted from a single beam (right) for SO, SiS, and $^{13}$CO. The vertical dashed line indicates the systemic velocity of 6.9~km~s$^{-1}$. For the individual spectrum, the RMS is computed as the standard deviation of line-free channels and is shown in the upper right of each panel. The shaded regions in the SO and SiS spectra show the velocity regions used to compute the column densities in Table \ref{tab:Ncol}. A non-Keplerian, compact emission counterpart to the proposed planet location is labeled in the $^{13}$CO spectrum.}
\label{fig:figure3}
\end{figure*}

In Figure \ref{fig:figure3}, we first extracted disk-integrated spectra for SO and SiS, as well as for $^{13}$CO to provide a reference for the Keplerian rotation of the disk. The SO emission is consistent with Keplerian rotation and has a similar double-peaked profile to that of $^{13}$CO. However, there is conspicuous excess flux at v$_{\rm{LSR}}{\approx}$8~km~s$^{-1}$, which represents the contribution of the bright NE blob. While the SO NE blob does not substantially deviate from Keplerian rotation, it has no bright localized counterpart on the other side of the disk, i.e., v$_{\rm{LSR}}<\rm{v}_{\rm{sys}}$. In contrast to SO, the SiS emission is entirely non-Keplerian with most emission coming from a narrow velocity range of v$_{\rm{LSR}}{\approx}$1--2.5~km~s$^{-1}$, while some fainter emission is present at v$_{\rm{LSR}}>$2.5~km~s$^{-1}$. This deviation from Keplerian rotation is clearly see in the channel maps (Figure \ref{fig:figure_appendix_sis}). The hand-drawn SiS masks are intentionally restrictive and the mask edges at v$_{\rm{LSR}}$=7.1~km~s$^{-1}$ are shown in Figure \ref{fig:figure3}. While the true SiS integrated flux may be larger than what is reported in Table \ref{tab:image_info}, we opted for this conservative approach to ensure high confidence in the SiS emission we consider.

Next, we extracted individual spectra within one beam size at the location of the bright NE blob in SO and SiS. The extraction positions are not identical given the small spatial offset (a few au) between the SO and SiS features (Figure \ref{fig:figure1}). As before, we also extracted a $^{13}$CO spectrum at the same position for reference. The SO emission associated with the SO NE blob is consistent with the Keplerian rotation of the disk, while the SiS emission shows a blueshift of 6~km~s$^{-1}$ from the expected Keplerian velocity as traced by $^{13}$CO. This SiS shift may point to a outflow around an embedded planet and we return to this in the Discussion. Intriguingly, we also note the presence of excess, non-Keplerian $^{13}$CO emission at a similar velocity (v$_{\rm{LSR}}{\approx}$5~km~s$^{-1}$) as the SiS emission.

More detailed analysis of the velocity structure of these lines is precluded by the relatively coarse channel spacing (${\approx}$0.4~km~s$^{-1}$) of the ALMA archival data. 

\subsection{Point-Source-Like Emission in $^{12}$CO and $^{13}$CO} \label{sec:CPD_in_CO}

\begin{figure*}
\centering
\includegraphics[width=\linewidth]{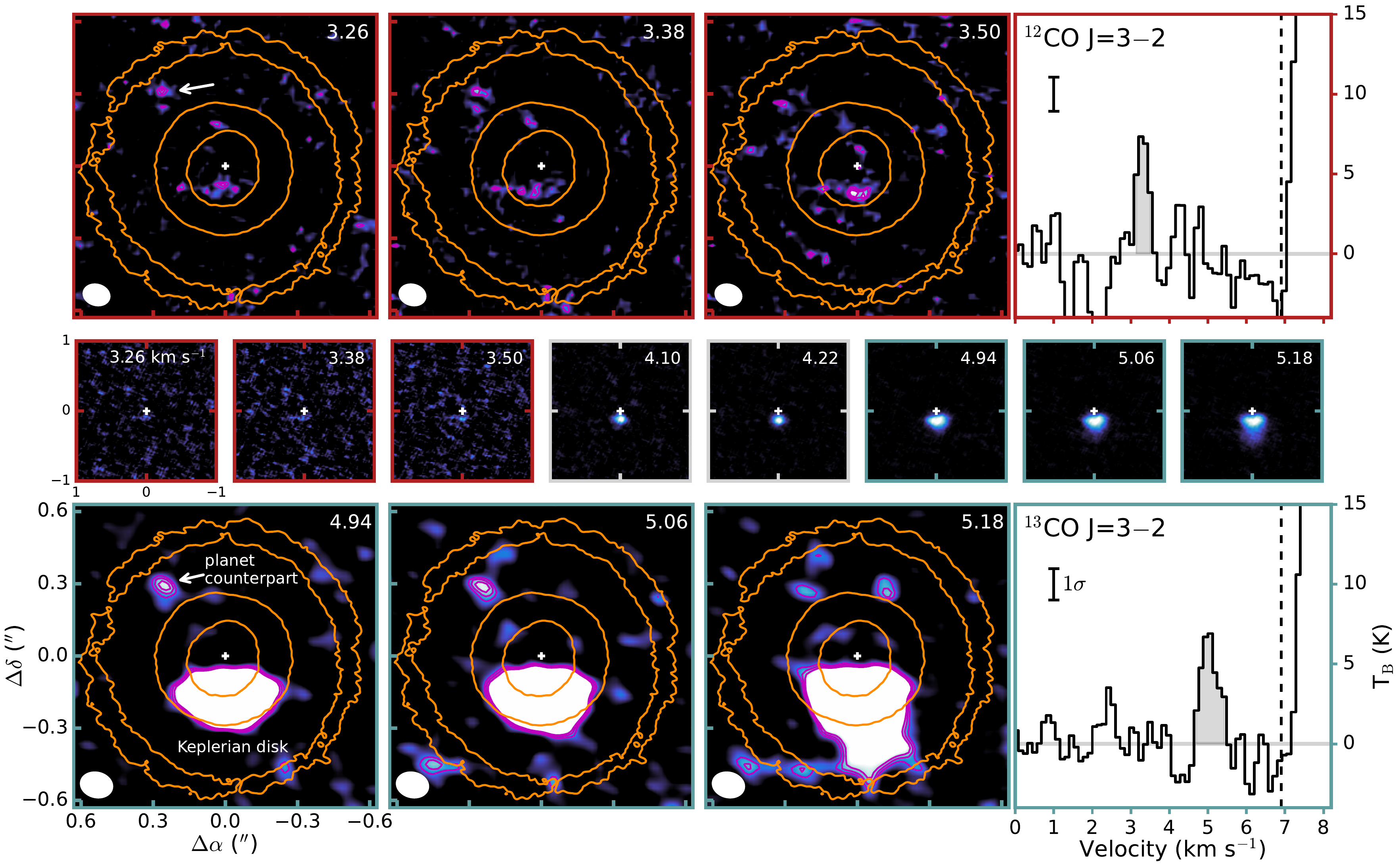}
\caption{Selected channel maps and spectra extracted from a single beam at the position of the point-source-like emission in $^{12}$CO J=3--2 (upper row; red border) and $^{13}$CO J=3--2 (lower row; blue border). The middle row shows the $^{12}$CO channels for reference with the border color corresponding to each set of channels. Several channels (gray border) between 3.50 and 4.94~km~s$^{-1}$ have been omitted for the sake of visual clarity. The orange contours show the millimeter dust rings. The colorscale shows the line emission and has been intentionally saturated to highlight the faint emission associated with a point source at a projected distance of 340-400~mas and PA${\approx}$38$^{\circ}$. Contours in pink show RMS$\times$[2.5, 3, 3.5]. The synthesized beam is shown in the lower left corner of each panel. We show the JvM-uncorrected images here to provide the most conservative estimate on the SNR of the compact $^{12}$CO and $^{13}$CO emission. A comprehensive analysis of both JvM- and non-JvM-corrected images are provided in Appendix \ref{sec:appendix_image_tests}.}
\label{fig:figure_CO}
\end{figure*}

As shown in Figure \ref{fig:figure_CO}, we also detected point-source-like emission in $^{12}$CO and $^{13}$CO at a projected distance of 340-400~mas (39-46~au) and PA${\approx}$38$^{\circ}$, which is near the location of the SO and SiS NE Blobs and the putative planet location. This emission is not related to the Keplerian component of the disk and no counterpart to either feature in $^{12}$CO or $^{13}$CO is seen on the other side of the disk (e.g., see full spectrum of $^{13}$CO in Figure \ref{fig:figure3}). We detected 3$\sigma$ emission in at least two adjacent channels for both $^{12}$CO and $^{13}$CO (Figure \ref{fig:figure_CO}). We confirmed that these features are robustly detected in both the JvM and non-JvM-corrected images, as well as in images generated from various combinations of \texttt{robust} parameters (see Appendix \ref{sec:appendix_image_tests} for full details).

The $^{12}$CO and $^{13}$CO emission do not share the same velocity, with the $^{12}$CO emission being more blueshifted (v$_{\rm{LSR}}{\approx}$3~km~s$^{-1}$) than the $^{13}$CO component (v$_{\rm{LSR}}{\approx}$5~km~s$^{-1}$). While the $^{12}$CO and $^{13}$CO emission are approximately spatially coincident, the $^{12}$CO emission appears slightly offset from the peak of $^{13}$CO (see Figure \ref{fig:planet_summary}). The compact $^{13}$CO is spatially-unresolved, while there is a slight hint that $^{12}$CO has a brighter northern component (see Figure \ref{fig:centroids}) with the caveat that the entire extent of the $^{12}$CO emission is within one beam. To better quantify this offset, we measured the centroids of the $^{12}$CO and $^{13}$CO emission in each channel with a full listing of the centroid locations in Appendix \ref{sec:CPD_Centroids}.

No counterpart in millimeter continuum emission is detected here or reported in previous high sensitivity continuum observations of this disk \citep[e.g.,][]{Perez19}. NIR colors of a point-source-like object suggests the presence of a significant amount of small dust at this location \citep{Hammond23}. This is consistent with the presence of a Jupiter-mass planet, as the lack of mm dust can be explained by strong dust filtration \citep{Rice06}, while the smaller dust grains remain coupled to the gas close to the protoplanet.

\subsection{Column Densities} \label{sec:Ncol}

Next, we used the measured line intensities of SO, SiS, and SO$_2$ to compute their column densities (for a detailed analysis of the CO gas density, see \citet{Garg22, Leemker22}). We followed a similar analysis as in \citet{Booth23} and assumed that the lines are optically thin and in local thermodynamic equilibrium (LTE). For SO and SiS, we only have one transition (and thus one upper state energy) and were unable to drive a rotational temperature (T$_{\rm{rot}}$). Likewise, for SO$_2$, both lines are undetected, which precludes a meaningful rotational diagram analysis. Instead, we adopted a constant rotational temperature of 100~K for all molecules. This is a reasonable assumption as all SO and SiS emission is confined to the inner disk (${<}$70~au) and is consistent with the warm gas temperatures reported in \citet{Leemker22}. 

\begin{figure}
\centering
\includegraphics[width=\linewidth]{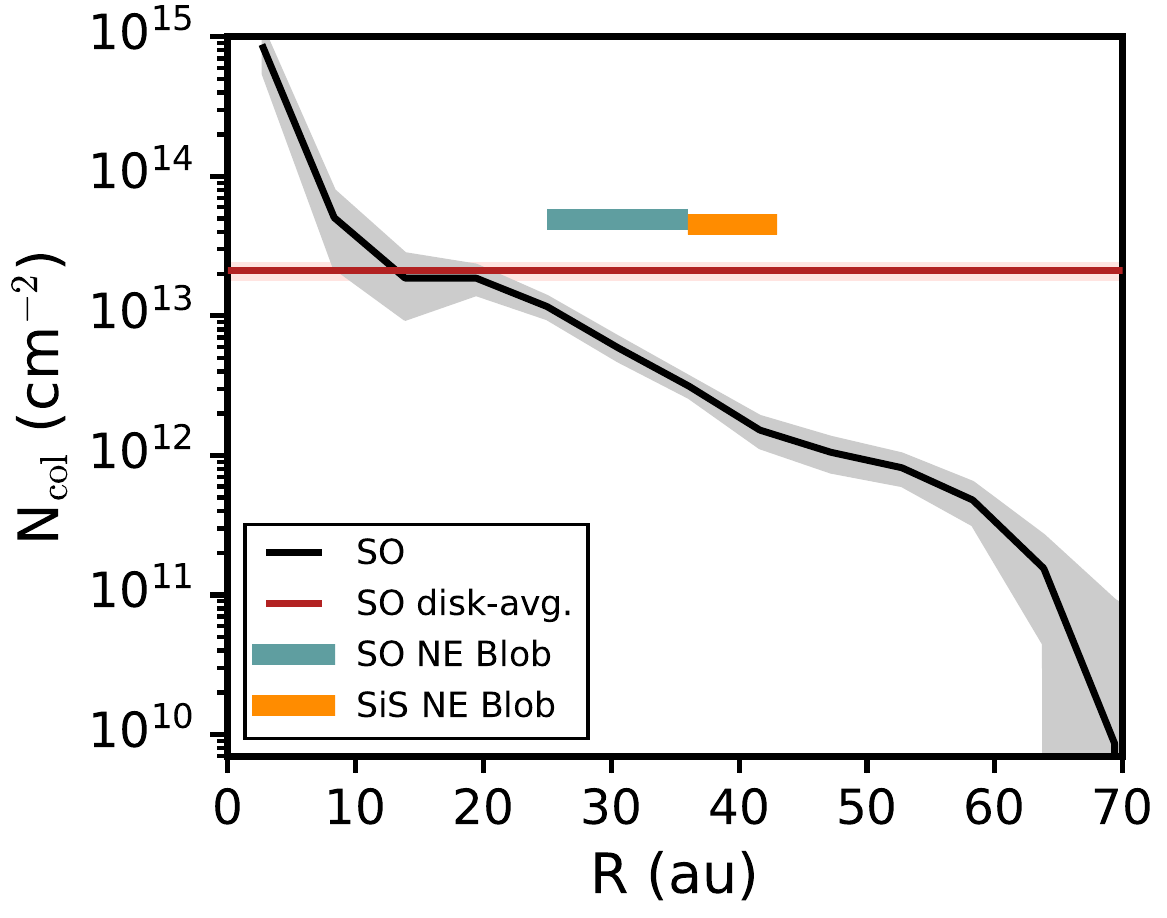}
\caption{Radial column densities of SO and SiS derived with a constant T$_{\rm{rot}}$=100~K. The SO radially-resolved profile is shown in black and the disk-averaged SO value is marked by a horizontal red line. The column densities of the NE Blob for SO and SiS are shown as teal and orange rectangles, respectively.}
\label{fig:figure5}
\end{figure}

Figure \ref{fig:figure5} shows the SO radial column density profile computed using the azimuthally-averaged radial line intensity profile (Figure \ref{fig:figure2}). Disk-averaged column densities for SO, SiS, and SO$_2$ were estimated from the integrated fluxes and upper limits reported in Table \ref{tab:image_info} and an assumed elliptical emitting area with the same position angle and inclination as the disk and a semi-major axis of 0\farcs6. Table \ref{tab:Ncol} lists the derived values.

We also derived the column densities at the location of the NE Blob. We used the fluxes of SO and SiS  (and 3$\sigma$ upper limit for SO$_2$) from spectra extracted within a single beam centered on the NE Blob position (Figure \ref{fig:figure3}). The SO column density is a factor of two larger than the disk-averaged values, while the SiS column density is nearly an order of magnitude larger, resulting in a nearly one-to-one SiS/SO ratio at this position.

It is possible that the gas temperatures at the NE Blob are considerably higher than 100~K, especially since SiS is observed in the gas phase, which suggests the presence of shocks and dust grain sputtering. We note that the column density calculations are only modestly sensitive to the assumed temperature, i.e., if the temperature were a factor of five larger, this results in a factor of two increase in SiS column density. Additional lines of SO and SiS with a sufficiently large range of upper state energies are required to better constrain the excitation conditions of both molecules.

\begin{deluxetable}{lccc}
\tablecaption{Derived Column Densities \label{tab:Ncol}}
\tablewidth{0pt}
\tablehead{
\colhead{Molecule} & \colhead{N$_{\rm{col}}$ (cm$^{-2}$)} & \colhead{ / SO (\%) }
}
\startdata
\textbf{Disk-Averaged} & \\
SO  & 2.1 $\pm$ 0.3 $\times$ 10$^{13}$ & \ldots \\
SiS & 6.5 $\pm$ 2.4 $\times$ 10$^{12}$ & 31 \\
SO$_2$ & ${<}$4.2 $\times$ 10$^{13}$ & ${<}$199\\ \\
\textbf{NE Blob} & \\
SO  & 4.9 $\pm$ 0.7 $\times$ 10$^{13}$ & \ldots  \\
SiS  & 4.5 $\pm$ 0.7 $\times$ 10$^{13}$ & 92 \\
SO$_2$ & ${<}$4.3 $\times$ 10$^{13}$ & ${<}$87 \\
\enddata
\tablecomments{All column densities are computed assuming a constant T$_{\rm{ex}}$=100~K. For the two non-detected SO$_2$ lines, we computed the upper limit as $3\sigma$ for both lines and report the smaller of the two.}
\end{deluxetable} \vspace{-12pt}

\section{Discussion} \label{sec:discussion}

In Section \ref{sec:chemical_sign}, we discuss the nature and origins of the compact emission observed in multiple molecular tracers in the context of ongoing giant planet formation in the HD~169142 disk. In Section \ref{sec:origins_of_SO}, we then explore the chemical origins of the broader ring-like SO emission, including the prominent asymmetry of the inner SO ring and compare our observations to other Class~II disks.

\subsection{Chemical Signatures of Ongoing Giant Planet Formation} \label{sec:chemical_sign}

\subsubsection{Compact $^{12}$CO and $^{13}$CO Emission Counterparts} \label{sec:CO_CPD}

The compact $^{12}$CO and $^{13}$CO emission is located at the center of a gas-depleted, mm and NIR annular gap \citep[e.g.,][]{Fedele17, Ligi18, Perez19, Garg22}; is coincident with the position of a high-intensity, Keplerian-rotating NIR point source \citep{Gratton19, Hammond23}; and lies along the same azimuthal arc as a known $^{12}$CO kinematic excess \citep{Garg22}. Figure \ref{fig:planet_summary} shows the close proximity of this compact emission with the location of the proposed planet from \citet{Gratton19, Hammond23}. These different lines of evidence suggest that we are observing molecular line emission associated with a giant planet embedded within the HD~169142 disk. This represents the third such feature seen in molecular gas after those identified in the AS~209 \citep{Bae21} and Elias 2-24 disks \citep{Pinte23} and the first seen in more than one molecular line.

With the current data, we cannot definitively determine the origin of this compact $^{12}$CO and $^{13}$CO emission. However, the significant blueshifts of both lines with respect to the systemic velocity suggest that this emission is not directly associated with a CPD or nearby circumplanetary material \citep{Perez15_gas_planets}. Instead, the observed velocities are more consistent with an outflow origin, where we are observing gas accelerating along our line of sight, while the velocity difference between the more blueshifted $^{12}$CO emission versus that of $^{13}$CO is likely an optical depth effect. The tentative spatial offset of $^{12}$CO from the more point-source-like emission of $^{13}$CO is also consistent with such a scenario.

Further detailed analysis is limited by the current SNR. Higher angular and spectral resolution data are required to better constrain the origin and properties of this compact $^{12}$CO and $^{13}$CO emission.

\subsubsection{Localized SO and Arc-like SiS Emission} \label{sec:SO_SiS_Discu}

We first discuss the potential chemical origins of SO and SiS, which are generally considered tracers of shocked gas, and then interpret the observed chemical signatures in the HD~169142 disk in the context of giant planet formation. Here, we only consider the compact SO that is co-spatial with the planet location. Please see Section \ref{sec:origins_of_SO} for a discussion of the bulk SO emission.

\textit{SO.} SO is often observed in regions of warm and shocked gas, including accretion shocks \citep{Sakai14, Sakai17, Oya19}, protostellar outflows \citep{Codella14, Taquet20}, MHD-driven disk winds \citep{Tabone17, Lee18}, and the warm inner envelopes of Class~I protostars \citep{Harsono21}. In shocks, the gas temperature rapidly increases to ${>}$100~K \citep{Draine83}, which enables efficient gas-phase formation of SO \citep[e.g.,][]{Prasad80, Hartquist80, vanGelder21} as well as the thermal desorption of any S-rich ices that are present \citep[e.g.,][]{Cleeves15, Kama19}. SO has been detected in a only handful of Class~II disks to date with evidence of gas-phase SO tracing the location of an embedded giant planet in at least one disk \citep{Booth23}. 

\textit{SiS.} SiS has been detected in a variety of settings in the ISM, including in the circumstellar envelopes of AGB stars \citep{Velilla19,Danilovich19}, massive star-forming regions \citep{Tercero11}, shocks around low-mass protostars \citep{Podio17}, and the innermost disks of massive protostars \citep{Tanaka20, Ginsburg23}. Unlike the more commonly-observed SiO molecule, which is a well-established tracer of silicates released from dust grain in shocks \citep[e.g.,][]{Gusdorf08}, the formation and destruction pathways of SiS remain less clear. While it is possible to produce gas-phase SiS through a direct release from dust cores, this requires strong shocks that fully destroy the grains and is inconsistent with existing observations of the low-mass protostellar shock L1157-B1, where SiS is not seen at the jet impact site, the location of the strongest shocks in this system \citep{Podio17}. Instead, SiS is thought to be a product of neutral-neutral gas-phase reactions between species released from dust grains in the shock \citep[e.g.,][]{Rosi18,Paiva20}. Recent theoretical work has shown a formation route of Si$+$SH that is efficient at warm temperatures (${\approx}$200~K) \citep{Mota21}, which are consistent with gas temperatures associated with an embedded giant plant. \citet{Zanchet18} have also suggested a formation route of Si$+$SO, which may be relevant in HD~169142 given the detection of SO. 

The observed chemical signatures of SO and SiS in the HD~169142 disk could thus originate from a variety of mechanisms, including a planet-driven outflow, circumstellar disk winds, planet-disk interactions driving an infall streamer, or emission directly from a circumplanetary disk/envelope.

A planet-driven outflow is the most consistent with the spatial and kinematic properties of both SiS and SO. The SiS has a substantially blueshifted velocity offset (Figure \ref{fig:figure3}) and extends from the planet location along an arc (Figure \ref{fig:figure1}), which taken together, suggest an origin in a localized outflow from an accreting protoplanet. This extended tail is co-spatial with the $^{12}$CO kinematic excess reported in \citet{Garg22} (see Figure \ref{fig:SiS_vs_CO_kinematcs}) and is located ahead in azimuth of the (clockwise) planet rotation. This is perhaps somewhat surprising, but we note that the geometry of the emission is not well-constrained and its apparent morphology may be influenced by projection effects. Given the observed blueshift of SiS and near-face on orientation of the HD~169142 disk, at least a substantial component of this outflow must be directed along the observer line of sight. Such a geometry is also consistent with an outflow origin of the $^{12}$CO and $^{13}$CO emission as discussed in Section \ref{sec:CO_CPD}.

The SO emission is considerably more compact than that of SiS, but it is difficult to tell if this is due to a lack of SO gas or excitation effects. The J=19--18 line of SiS has an upper state energy (E$_{\rm{u}}$=166~K) approximately twice that of SO J=8$_8$--7$_7$ (E$_{\rm{u}}$=88~K). However, with the current coarse velocity resolution and lack of information about SO excitation, we cannot rule out the presence of hot SO gas at or or near the location of the extended SiS emission. The SO emission also does not show the same blueshifted velocity offset as SiS. While this may suggest a non-outflow-based origin such as accretion shocks onto a CPD or a nearby warm gas envelope, it is possible that SO is tracing a different, lower-velocity component of the same outflow. 

\begin{figure}[!]
\centering
\includegraphics[width=\linewidth]{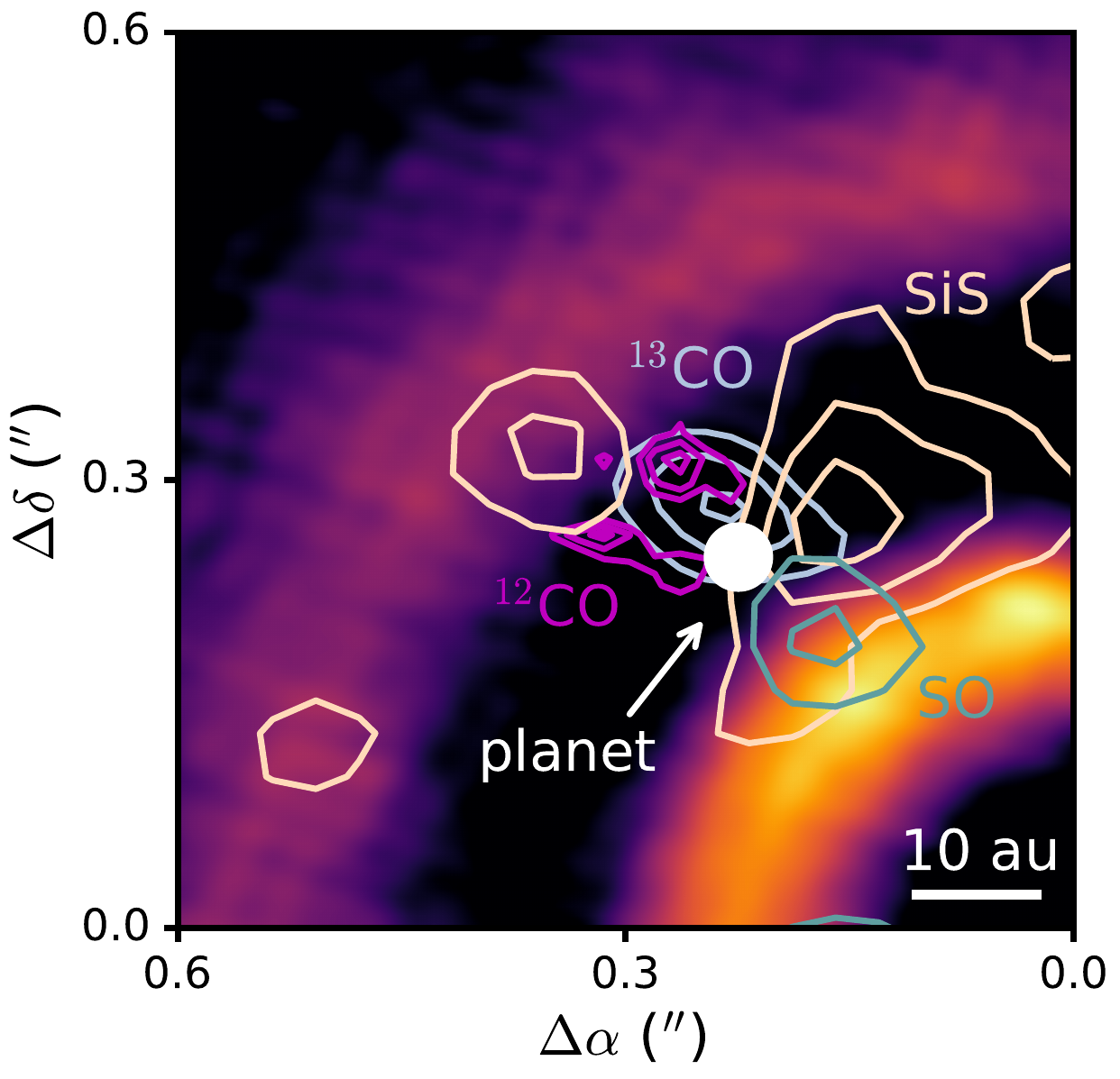}
\caption{A gallery of chemical signatures related to the proposed planet in the HD~169142 disk. The peak intensities of $^{12}$CO (pink), $^{13}$CO (blue), SiS (peach) and SO (teal) are shown in contours and the millimeter dust is in the colorscale. The planet location is indicated by a white circle \citep{Gratton19, Hammond23}. A scale bar showing 10~au is in the lower right corner.}
\label{fig:planet_summary}
\end{figure}

While there are several alternate explanations for these chemical signatures, they are generally insufficient to simultaneously explain the emission morphology and kinematics of both SiS and SO. Disk winds appear inconsistent with the localized nature of the SO and SiS emission and relatively small velocity deviations observed (disk winds typically show velocity offsets of tens of km~s$^{-1}$ from Keplerian rotation, e.g., \citealp{Booth21_MAPS}). Alternatively, simulations have shown that gap edges become mildly Rayleigh unstable and intermittently shed streams of material into the gap in the presence of high planet masses \citep{Fung16}. However, it is not clear if such wakes result in sufficiently strong shocks to produce detectable amounts of gas-phase SiS. Moreover, detailed analysis of CO kinematics in the HD~169142 disk find no evidence for such planetary wakes \citep{Garg22}. While NIR spirals arms are observed in HD~169142, neither SiS or SO traces the locations of these spirals, which are located behind the planet orbit in azimuth \citep{Hammond23}. It is also unlikely that we are seeing emission from the circumplanetary material itself. The extended morphology of SO, and especially of SiS, along with mutual offsets of a few au from the planet location and CO counterparts (Figure \ref{fig:planet_summary}), are inconsistent with SO and SiS directly originating from close-in circumplanetary material. 

Overall, this suggests that the most plausible scenario is one in which an embedded, giant planet identified in both NIR and via compact $^{12}$CO and $^{13}$CO emission is driving an outflow, which locally heats gas and produces an extended shock traced in SO and SiS.

\subsection{Morphology and Origins of SO Emission} \label{sec:origins_of_SO}

\subsubsection{Chemical Origins and Asymmetry} \label{sec:SO_asym}

The SO emission distribution in the HD~169142 disk is confined to the innermost ${\approx}$70~au and closely follows that of the millimeter dust (Figure \ref{fig:figure1}). The bright asymmetric inner ring is located at the edge of the inner dust cavity, while the faint outer ring appears to peak at the edge of the outer dust ring. This emission distribution is broadly consistent with SO originating from the thermal desorption of S-rich ices \citep[e.g.,][]{Cleeves11, Kama19}. This may be due to the direct sublimation of SO ice or gas-phase chemistry following the UV photo-dissociation of evaporated H$_2$S and H$_2$O ices. In the later case, SO can form via the efficient barrierless gas-phase neutral-neutral reactions, S $+$ OH or O $+$ SH \citep[e.g.,][]{Charnley97}. Since both OH and H$_2$O were not detected with \textit{Herschel}/PACS (The Photodetector Array Camera and Spectrometer) \citep{Fedele13} in this disk, this suggests that the ices in HD~169142 are H$_2$S-rich, although other forms of S-bearing ices cannot be ruled out (e.g., OCS, SO$_2$). 

If the ice reservoir is constrained to the larger mm- and cm-sized dust grains, then the observed asymmetry in SO is difficult to explain with ice sublimation alone since the dust is largely axisymmetric \citep{Fedele17, Macias19, Perez19}. However, SO asymmetries can be caused by changing physical conditions in the disk, namely azimuthal temperature variations due to a warped disk or dynamical interactions with an embedded planet. We explore each of these scenarios in turn.

\textit{Misaligned Inner Disk.} Asymmetries in molecular line emission can arise in warped disks from azimuthal variations in disk illumination by the central star, which in turn manifests in chemical variations. Warped disks do not necessarily show clear azimuthal asymmetries in their outer disks, but instead possess misaligned inner disks due to an embedded companion \citep{Young21}. Models have identified SO as a potential chemical tracer of warped disks and demonstrated that changing X-ray illumination drives variations in SO abundance \citep{Young21}. The HD~169142 disk shows several signatures of a misaligned inner disk, including shadowing in NIR polarized intensity images \citep{Quanz13, Pohl17, Bertrang18, Rich22}. On sub-au scales, observations with the GRAVITY instrument at the Very Large Telescope revealed a considerable misalignment in both inclination (i$_{\rm{in}}$=35$^{\circ}$) and position angle (PA$_{\rm{in}}$=32$^{\circ}$) versus the outer disk (i=13$^{\circ}$, PA=5$^{\circ}$) as traced by CO \citep{Bohn22}. \citet{Garg22} suggested that the kinematic excess observed in $^{12}$CO may also originate from a misaligned inner disk, as spurious kinematic residuals can be generated from the subtraction of Keplerian models that incorrectly assume a constant PA and inclination \citep{Young22}. The HD~169142 disk also shows evidence for a giant planet (${\approx}$1-10~M$_{\rm{Jup}}$) in its inner dust cavity in observations in the NIR \citep{Bertrang20} and of CO isotopologues \citep{Leemker22}. The models of SO chemical asymmetries in warped disks from \citet{Young21} were based on a hydrodynamical simulation of a 6.5~M$_{\rm{Jup}}$ planet embedded in a disk at a radius of 5~au around a solar-mass star orbiting at a 12$^{\circ}$ inclination with respect to the disk \citep{Nealon19}. Thus, to first order, this is similar to the HD~169142 system and implies that the presence of a warped disk, due to an embedded giant planet and misaligned inner disk, offers a plausible explanation for the observed SO brightness asymmetry.

\textit{Planet-disk Interactions.} It is also possible for dynamical interactions between forming planets and the circumstellar disk to produce chemical asymmetries \citep[e.g.,][]{Cleeves15}. We see evidence of local gas heating at the proposed planet location in the form of compact SO emission (see Section \ref{sec:SO_SiS_Discu}). This additional heating may also contribute, at least in part, to the broader SO asymmetry. The asymmetric, ring-like SO emission likely does not trace accretion shocks but rather the impact of this additional heating on the surrounding disk. It has been shown, for instance, that an embedded giant planet can excite the orbits of planetesimals within in the circumstellar gas disk and cause bow-shock heating that evaporates ices \citep{Nagasawa19}. While the peak of the SO emission brightness is offset in PA from the planet location (Figure \ref{fig:figure1}), we cannot currently discern if this offset has a dynamical origin or is an excitation effect, i.e., we are not observing the hottest SO gas near to the planet location.

\subsubsection{Comparison to Other Disks} \label{sec:compare_other_disks}

While SO is commonly observed in protostellar systems \citep[e.g.,][]{Bachiller97, Codella14, Taquet20, Garufi22}, detections are relatively rare in evolved (${>}$1~Myr) Class II protoplanetary disks. HD~169142 is only the fifth such disk where SO has been spatially resolved, after AB~Aur \citep{Pacheco16,Riviere20}, Oph-IRS~48 \citep{Booth21_IRS}, HD~100546 \citep{Booth23}, and DR~Tau \citep{Huang23}. All of these sources are transition disks around Herbig stars, with the exception of DR~Tau, which shows substantial interaction with its environment. There is also a marginal detection of SO in single dish observations of the transition disk around the T~Tauri star GM~Aur \citep{Guilloteau16}, but this disk also shows evidence of late infall from a remnant envelope or cloud material \citep{Huang21}. Hence, it is not known where the SO is originating from, i.e., disk vs. shocks from cloud material. Taken together, this points to ice sublimation as a common origin for the bulk of SO emission in isolated or non-interacting Class~II disks. 

The HD~169142 disk appears typical in its SO content relative to other Herbig disks. Its disk-integrated column density is consistent within a factor of a few compared to the HD~100546 (4.0--6.4$\times10^{13}$~cm$^{-2}$) \citep{Booth23} and AB~Aur disks (2.1--3.4$\times10^{13}$~cm$^{-2}$) \citep{Riviere20}. The Oph-IRS~48 disk has a considerably larger SO column density (4.7--9.9$\times10^{15}$~cm$^{-2}$) \citep{Booth21_IRS}, but this disk displays several unique properties, including a high degree of chemical complexity not seen in other Class~II disks \citep{vanderMarel21, Brunken22}, which includes the only detection of SO$_2$ in disks to date \citep{Booth21_IRS}.

Asymmetries are a common feature of SO emission in protoplanetary disks with all spatially-resolved SO observations showing some degree of azimuthal asymmetry. Of those disks with spatially-resolved SO, three sources have confirmed \citep[AB Aur~b,][]{Currie22} or proposed embedded planets (\citealp[HD~100546,][]{Currie15, Brittain19}; \citealp[HD~169142,][]{Gratton19}). The AB~Aur disk hosts a particularly complex and dynamic environment, with spiral arms, high levels of accretion and outflow activity and is possibly gravitationally unstable \citep{Tang12, Salyk13, Riviere20,Cadman21}. Thus, disentangling the origin of the SO asymmetry in such a setting is challenging. The HD~100546 disk, however, is a much closer analogue to the HD~169412 system, as both are warm, Herbig disks with compelling evidence of embedded planets. In both cases, there is a component of SO emission that traces the location of proposed giant planets. In the HD~100546 disk, \citet{Booth23} also detected temporal variability in the SO emission, which was speculated to be in response to the orbit of an embedded planet. It may be possible to observe similar variability in SO (and possibly SiS) in future observations of the HD~169142 system, although this will require a larger observation baseline due to the longer rotation period (${\approx}$174~yr) associated with a planet at 38~au compared to a 10~au planet in the HD~100546 disk. We do, however, note that these disks are not perfect analogues, as SiS is not detected in HD~100546 and has a SiS/SO column density ratio of ${<}$5.0\% \citep{Booth23}, while this ratio is nearly one-to-one in HD~169412 (Table \ref{tab:Ncol}).

\section{Conclusions} \label{sec:conlcusions}

The HD~169142 disk shows several distinct chemical signatures that demonstrate a compelling link to ongoing giant planet formation. Using ALMA archival data, we identify compact SO J=8$_8$--7$_7$ and SiS J=19--18 emission coincident with the position of a proposed ${\sim}$2~M$_{\rm{Jup}}$ planet seen as a localized, Keplerian NIR feature within a gas-depleted, annular dust gap at ${\approx}$38~au. The SiS emission is non-Keplerian and is located along the same extended azimuthal arc as a known $^{12}$CO kinematic excess. This is the first tentative detection of SiS emission in a protoplanetary disk and suggests that the planet is driving sufficiently strong shocks to produce gas-phase SiS. We also report the discovery of compact $^{12}$CO J=3--2 and $^{13}$CO J=3--2 emission counterparts coincident with the planet location. Taken together, a planet-driven outflow provides the most consistent explanation for the spatial and kinematic properties of these chemical signatures and their close proximity to the planet position.

In addition to the localized SO emission, we resolve a bright, azimuthally-asymmetric SO ring at ${\approx}$24~au located at the inner edge of the central dust cavity. This makes HD~169142 only the fifth such Class~II system with a spatially-resolved SO detection. The bulk of the SO emission likely has an origin in ice sublimation, while its asymmetric distribution suggests the presence of azimuthal temperature variations driven by a misaligned inner disk or planet-disk interactions, possibly due to the giant planet at ${\approx}$38~au.

The HD~169142 system presents a powerful template for future searches of planet-related chemical asymmetries in protoplanetary disks and an ideal test-bed for observational follow-up. The sudden increase in temperature and density in the shocked gas around the embedded planet likely results in a hot gas-phase chemistry, in which the abundance of several molecular species, in addition to Si- or S-bearing molecules, dramatically increases by several orders of magnitude. Deep spectral surveys of the location around the embedded planet may reveal new chemistry not yet detected in planet-forming disks and allow for new insights into planet-feeding gas. In parallel to this, a survey of SO and SiS emission in disks with known or suspected giant planets would provide a vital confirmation of these molecules as novel tracers of embedded planets. \\

The authors thank the anonymous referee for valuable comments that improved both the content and presentation of this work. We thank Sean Andrews and Richard Teague for useful discussions. This paper makes use of the following ALMA data: ADS/JAO.ALMA\#2012.1.00799.S and 2015.1.00806.S. ALMA is a partnership of ESO (representing its member states), NSF (USA) and NINS (Japan), together with NRC (Canada), MOST and ASIAA (Taiwan), and KASI (Republic of Korea), in cooperation with the Republic of Chile. The Joint ALMA Observatory is operated by ESO, AUI/NRAO and NAOJ. The National Radio Astronomy Observatory is a facility of the National Science Foundation operated under cooperative agreement by Associated Universities, Inc. K.I.\"O. acknowledges support from the Simons Foundation (SCOL \#686302) and the National Science Foundation under Grant No. AST-1907832. 

%

\facilities{ALMA}


\software{Astropy \citep{astropy_2013,astropy_2018}, \texttt{bettermoments} \citep{Teague18_bettermoments}, CASA \citep{McMullin_etal_2007}, \texttt{cmasher} \citep{vanderVelden20}, \texttt{GoFish} \citep{Teague19JOSS}, \texttt{keplerian\_mask} \citep{rich_teague_2020_4321137}, Matplotlib \citep{Hunter07}, NumPy \citep{vanderWalt_etal_2011}, \texttt{Photutils} \citep{larry_bradley_2022_6825092}}



\clearpage
\appendix

\section{Observational Details} \label{sec:observational details}

Table \ref{tab:full_obs_program_details} lists all ALMA execution blocks used in this work and includes the ALMA project codes, PIs, UT observing dates, number of antennas, on-source integration times, baseline ranges, observatory-estimated spatial resolutions, maximum recoverable scales (M.R.S.), mean precipitable water vapor (PWV), and flux, phase, and bandpass calibrators.

\begin{deluxetable*}{llcccccccccc}[!htpb]
\tablecaption{Details of Archival ALMA Observations \label{tab:full_obs_program_details}}
\tablewidth{0pt}
\tablehead{
\colhead{Project} & \colhead{PI} & \colhead{UT Date} & \colhead{No. Ants.} & \colhead{Int.} & \colhead{Baselines} & \colhead{Res.} & \colhead{M.R.S.} & \colhead{PWV} & \multicolumn3c{Calibrators} \\ \cline{10-12} 
\colhead{Code} & \colhead{} & \colhead{} & \colhead{} & \colhead{(min)} & \colhead{(m)} & \colhead{($^{\prime \prime}$)} & \colhead{($^{\prime \prime}$)}  & \colhead{(mm)} & \colhead{Flux} & \colhead{Phase} & \colhead{Bandpass}
}
\startdata
2012.1.00799.S & M. Honda & 2015-07-26 & 41 & 41.4 & 15--1574 & 0.13 & 1.3 & 0.4 & J1924-2914 & J1826-2924 & J1924-2914 \\
               &          & 2015-07-27 & 41 & 21.2 & 15--1574 & 0.13 &  1.4 & 0.3 & J1924-2914 & J1826-2924 & J1924-2914\\
               &          & 2015-08-08 & 43 & 41.4 & 35--1574 & 0.13 & 1.3 & 1.3 & Pallas & J1826-2924 & J1924-2914 \\               
2015.1.00806.S & \text{J. Pineda} & \text{2015-12-06} & 32 & 25.3 & 19--7716 & 0.031 & $0.39$ & $1.0$ & J1733-1304 & J1826-2924  & J1924-2914
\enddata
\end{deluxetable*}

\section{Properties of Compact Emission in SO, $^{12}$CO, and $^{13}$CO with Various Imaging Parameters} \label{sec:appendix_image_tests}
\begin{figure*}[p!]
\centering
\includegraphics[width=\linewidth]{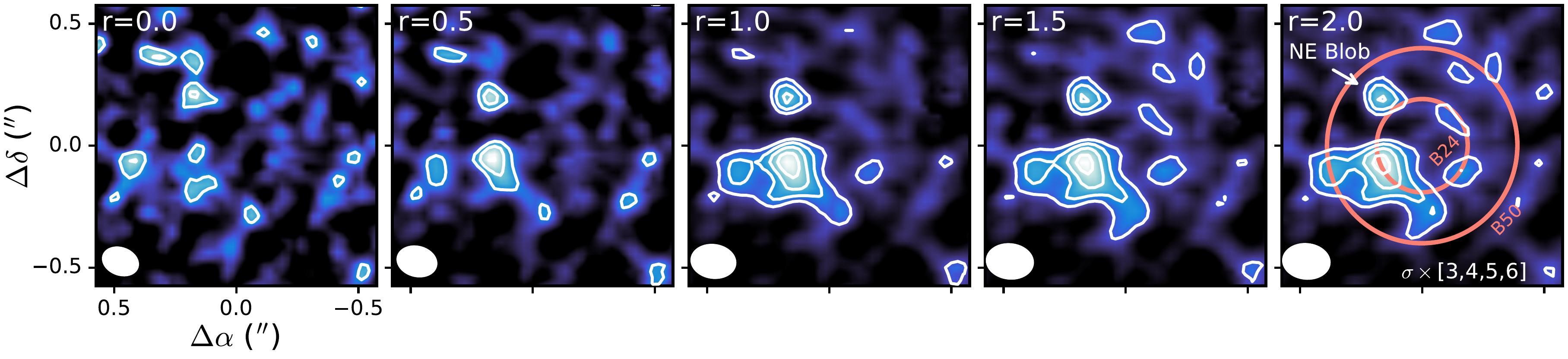}
\caption{Gallery of SO peak intensity maps generated from images with a range of Briggs \texttt{robust} values from 0.0 to 2.0. The location of the NE Blob and the two SO rings are labeled in the rightmost panel.  Contours show RMS$\times$[3,4,5,6]. The synthesized beam is shown in the lower left corner of each panel.}
\label{fig:figure_SO_image}
\end{figure*}

Figure \ref{fig:figure_SO_image} shows SO peak intensity maps for a range of Briggs \texttt{robust} values (0.0 to 2.0). The compact SO emission (NE Blob) is clearly detected in all images. Images with lower \texttt{robust} values lack sufficient sensitivity to detect either the NE Blob or the overall ring-like SO structure. Table \ref{tab:img_tests} shows a summary of the properties of the compact SO emission measured from each of these images. We computed the RMS, central velocity, FWHM, peak intensity, and integrated intensity within a synthesized beam around the NE Blob. We find that the NE Blob is well-detected with peak intensities with SNRs of ${\approx}$5-8.  As shown in the last panel of Figure \ref{fig:figure_SO_image}, it is also clear that the NE Blob is at a radius distinct from either broad SO ring at 24~au or 50~au and is located near the position of the HD~169142~b planet.

\begin{deluxetable*}{lcccccccc}
\tablecaption{Summary of Compact SO Emission Properties in Various Images \label{tab:img_tests}}
\tabletypesize{\normalsize} \tablewidth{0pt}
\tablehead{
\colhead{ \texttt{robust} } & \colhead{Beam} & \colhead{PA} & \colhead{RMS\tablenotemark{a}} & \colhead{Central Velocity} & \colhead{FWHM}  & \colhead{Peak Int.} & \colhead{Integrated Int.} \vspace{-0.1cm} \\
\colhead{} & \colhead{(mas $\times$ mas)} & \colhead{(deg)} & \colhead{(mJy beam$^{-1}$)} & \colhead{(km s$^{-1}$)} &  \colhead{(km s$^{-1}$)} & \colhead{(mJy beam$^{-1}$)} & \colhead{(mJy beam$^{-1}$ km s$^{-1}$})
}
\startdata
0.0 & 150~$\times$~111 & 69.4 & 1.41 & 8.2~$\pm$~0.1 & 1.9~$\pm$~0.1 & 7.2~$\pm~$0.7 & 11.7~$\pm$~1.2 &  \\
0.5 & 164~$\times$~123 & 76.0 & 1.06 & 7.9~$\pm$~0.1 & 1.4~$\pm$~0.1 & 6.7~$\pm~$0.7 & 9.6~$\pm$~1.0 &  \\
1.0 & 183~$\times$~137 & 82.2 & 0.87 & 7.9~$\pm$~0.1 & 1.5~$\pm$~0.1 & 7.0~$\pm~$0.7 & 9.4~$\pm$~0.9 &  \\
1.5 & 192~$\times$~143 & 84.1 & 0.88 & 8.0~$\pm$~0.1 & 1.5~$\pm$~0.1 & 6.5~$\pm~$0.6 & 9.3~$\pm$~0.9 &  \\
2.0 & 193~$\times$~144 & 84.4 & 0.89 & 8.0~$\pm$~0.1 & 1.6~$\pm$~0.1 & 6.7~$\pm~$0.7 & 9.8~$\pm$~1.0 &  \\
\enddata
\tablenotetext{a}{The RMS noise was computed within the synthesized beam around the compact SO emission over the first 20 line-free channels of the cube.}
\tablecomments{We adopt a 10\% systematic flux uncertainty and quarter-channel velocity uncertainties.}
\end{deluxetable*}

\begin{figure*}[p!]
\centering
\includegraphics[width=\linewidth]{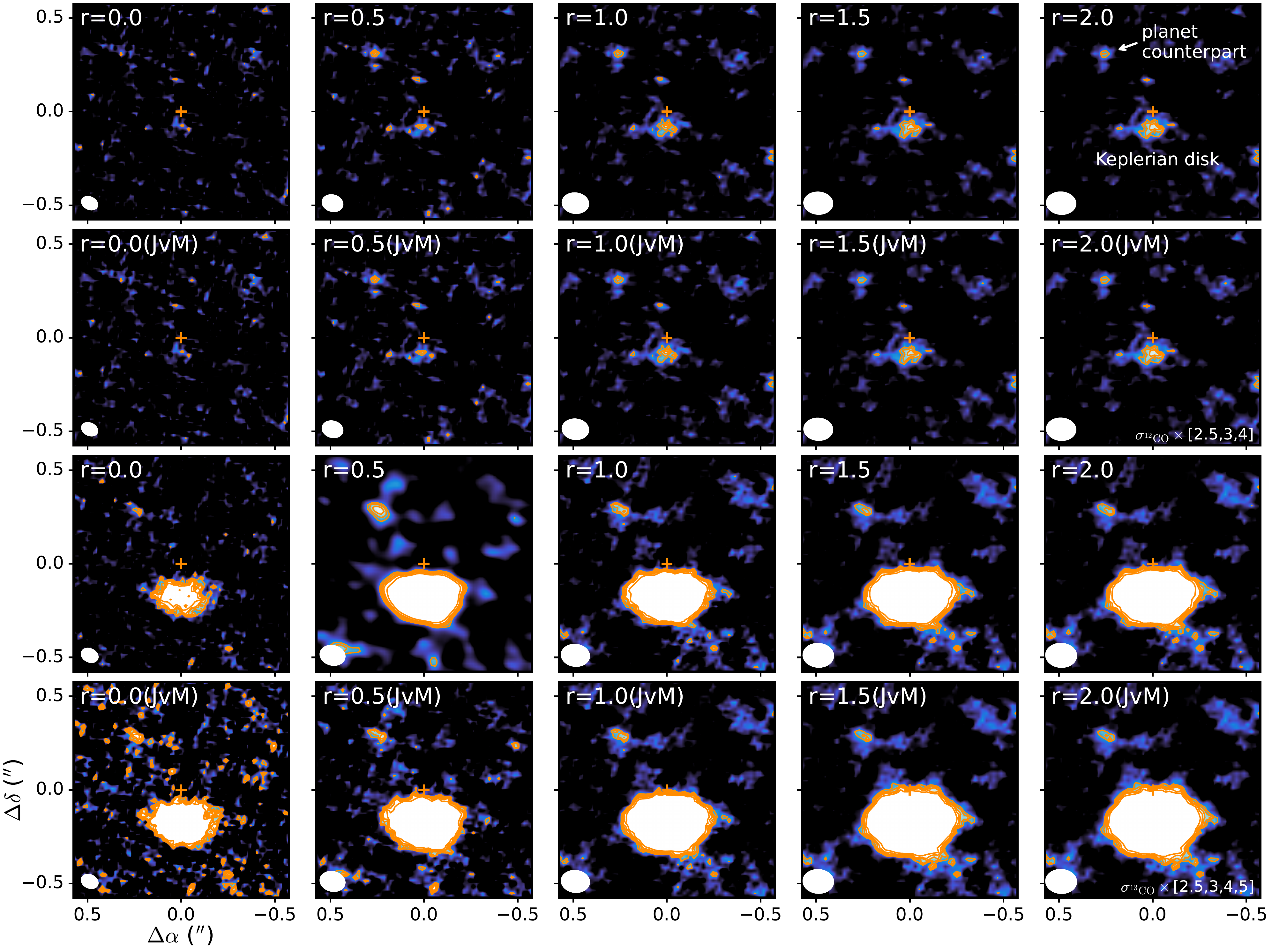}
\caption{Gallery of selected $^{12}$CO and $^{13}$CO channels (3.22 and 4.96~km~s$^{-1}$, respectively), which show the peak intensity of the emission counterparts. Images were generated with a range of Briggs \texttt{robust} values from 0.0 to 2.0 (left to right) and without (top) and with (bottom) the JvM correction. The locations of the compact emission counterparts and the expected Keplerian rotation are marked in the upper rightmost panel. Contours show RMS$\times$[2.5,3,4] and $\times$[2.5,3,4,5] for $^{12}$CO and $^{13}$CO, respectively. The disk center is marked by a `+' and the synthesized beam is shown in the lower left corner of each panel.}
\label{fig:12CO_13CO_img_tests}
\end{figure*}

Figure \ref{fig:12CO_13CO_img_tests} and Table \ref{tab:img_tests_CO} show a similar analysis of the properties of the point-source-like emission in $^{12}$CO and $^{13}$CO in a range of images with different Briggs \texttt{robust} values from 0.0 to 2.0. Images with lower \texttt{robust} values lacked sufficient sensitivity to detect the compact $^{12}$CO and $^{13}$CO emission. For each \texttt{robust} value, we also generated images both with and without the JvM-correction. Localized emission in the vicinity of HD~169142~b is detected in all image combinations with the exception of the $^{12}$CO image at a \texttt{robust} of 0.0. However, this image has a considerably higher RMS and the smallest beam size so this non-detection is likely due to insufficient sensitivity at these small beam sizes. Depending on the image, the point-source-like $^{12}$CO emission was detected with peak SNRs ranging from ${\approx}$3-4, while the $^{13}$CO emission has peak SNRs of ${\gtrsim}$3-7. Thus, the detection of $^{12}$CO and $^{13}$CO emission near HD~169142~b is statistically significant regardless of which image we use.

\begin{deluxetable*}{llccccccc}
\tablecaption{Summary of Point-source-like $^{12}$CO and $^{13}$CO Emission Properties in Various Images \label{tab:img_tests_CO}}
\tabletypesize{\normalsize} \tablewidth{0pt}
\tablehead{
\colhead{ \texttt{robust} } & \colhead{Beam} & \colhead{PA} & \colhead{RMS\tablenotemark{a}} & \colhead{Central Velocity} & \colhead{FWHM}  & \colhead{Peak Int.} & \colhead{Integrated Int.} \vspace{-0.1cm} \\
\colhead{} & \colhead{(mas $\times$ mas)} & \colhead{(deg)} & \colhead{(mJy beam$^{-1}$)} & \colhead{(km s$^{-1}$)} & \colhead{(km s$^{-1}$)} & \colhead{(mJy beam$^{-1}$)} & \colhead{(mJy beam$^{-1}$ km s$^{-1}$})
}
\startdata
\textbf{$^{12}$CO J=3$-$2}
 \\\textbf{no  JvM} \\
0.0 & 90~$\times$~66 & 63.5 & 3.03 & 3.33~$\pm$~0.03 & 0.30~$\pm$~0.03 & 7.6~$\pm$~0.8 & 2.0~$\pm$~0.2 &  \\
0.5 & 112~$\times$~88 & 74.1 & 2.34 & 3.31~$\pm$~0.03 & 0.36~$\pm$~0.03 & 7.1~$\pm$~0.7 & 2.5~$\pm$~0.2 &  \\
1.0 & 142~$\times$~109 & 83.4 & 1.51 & 3.34~$\pm$~0.03 & 0.43~$\pm$~0.03 & 5.6~$\pm$~0.6 & 1.7~$\pm$~0.2 &  \\
1.5 & 155~$\times$~118 & 86.1 & 1.21 & 3.36~$\pm$~0.03 & 0.48~$\pm$~0.03 & 4.5~$\pm$~0.4 & 1.4~$\pm$~0.1 &  \\
2.0 & 157~$\times$~119 & 85.6 & 1.18 & 3.36~$\pm$~0.03 & 0.48~$\pm$~0.03 & 4.4~$\pm$~0.4 & 1.4~$\pm$~0.1 &  \\
\hline\textbf{$^{12}$CO J=3$-$2}
 \\\textbf{with JvM} \\
0.0 & 90~$\times$~66 & 63.5 & 1.71 & 3.33~$\pm$~0.03 & 0.30~$\pm$~0.03 & 4.3~$\pm$~0.4 & 1.1~$\pm$~0.1 &  \\
0.5 & 112~$\times$~87 & 74.1 & 1.49 & 3.31~$\pm$~0.03 & 0.36~$\pm$~0.03 & 4.5~$\pm$~0.5 & 1.6~$\pm$~0.2 &  \\
1.0 & 142~$\times$~109 & 83.4 & 0.87 & 3.34~$\pm$~0.03 & 0.43~$\pm$~0.03 & 3.3~$\pm$~0.3 & 1.0~$\pm$~0.1 &  \\
1.5 & 154~$\times$~118 & 86.1 & 0.62 & 3.36~$\pm$~0.03 & 0.48~$\pm$~0.03 & 2.3~$\pm$~0.2 & 0.7~$\pm$~0.1 &  \\
2.0 & 157~$\times$~119 & 85.6 & 0.60 & 3.36~$\pm$~0.03 & 0.48~$\pm$~0.03 & 2.2~$\pm$~0.2 & 0.7~$\pm$~0.1 &  \\
\hline\textbf{$^{13}$CO J=3$-$2}
 \\\textbf{no  JvM} \\
0.0 & 94~$\times$~71 & 63.4 & 1.89 & 5.04~$\pm$~0.03 & 0.54~$\pm$~0.03 & 11.6~$\pm$~1.2 & 5.8~$\pm$~0.6 &  \\
0.5 & 135~$\times$~105 & 76.3 & 1.45 & 5.01~$\pm$~0.03 & 0.51~$\pm$~0.03 & 8.7~$\pm$~0.9 & 3.9~$\pm$~0.4 &  \\
1.0 & 151~$\times$~117 & 83.5 & 1.46 & 5.01~$\pm$~0.03 & 0.44~$\pm$~0.03 & 6.1~$\pm$~0.6 & 2.6~$\pm$~0.3 &  \\
1.5 & 163~$\times$~127 & 86.0 & 1.59 & 5.00~$\pm$~0.03 & 0.41~$\pm$~0.03 & 5.4~$\pm$~0.5 & 2.2~$\pm$~0.2 &  \\
2.0 & 165~$\times$~128 & 86.5 & 1.62 & 5.00~$\pm$~0.03 & 0.42~$\pm$~0.03 & 5.4~$\pm$~0.5 & 2.2~$\pm$~0.2 &  \\
\hline\textbf{$^{13}$CO J=3$-$2}
 \\\textbf{with JvM} \\
0.0 & 94~$\times$~71 & 63.4 & 1.08 & 5.04~$\pm$~0.03 & 0.54~$\pm$~0.03 & 6.6~$\pm$~0.7 & 3.3~$\pm$~0.3 &  \\
0.5 & 135~$\times$~105 & 76.3 & 0.88 & 5.01~$\pm$~0.03 & 0.51~$\pm$~0.03 & 6.0~$\pm$~0.6 & 2.7~$\pm$~0.3 &  \\
1.0 & 151~$\times$~117 & 83.5 & 0.85 & 5.01~$\pm$~0.03 & 0.44~$\pm$~0.03 & 3.6~$\pm$~0.4 & 1.5~$\pm$~0.2 &  \\
1.5 & 163~$\times$~127 & 86.0 & 0.85 & 5.00~$\pm$~0.03 & 0.41~$\pm$~0.03 & 2.9~$\pm$~0.3 & 1.2~$\pm$~0.1 &  \\
2.0 & 165~$\times$~128 & 86.5 & 0.85 & 5.00~$\pm$~0.03 & 0.42~$\pm$~0.03 & 2.8~$\pm$~0.3 & 1.2~$\pm$~0.1 &  \\
\hline\enddata
\tablenotetext{a}{The RMS noise was computed within the synthesized beam around the compact $^{12}$CO and $^{13}$CO emission over the first 20 line-free channels of the cube.}
\tablecomments{We adopt a 10\% systematic flux uncertainty and quarter-channel velocity uncertainties.}
\end{deluxetable*}

\section{Channel Maps} \label{sec:appendix_channel_maps}

\begin{figure*}[p!]
\centering
\includegraphics[width=\linewidth]{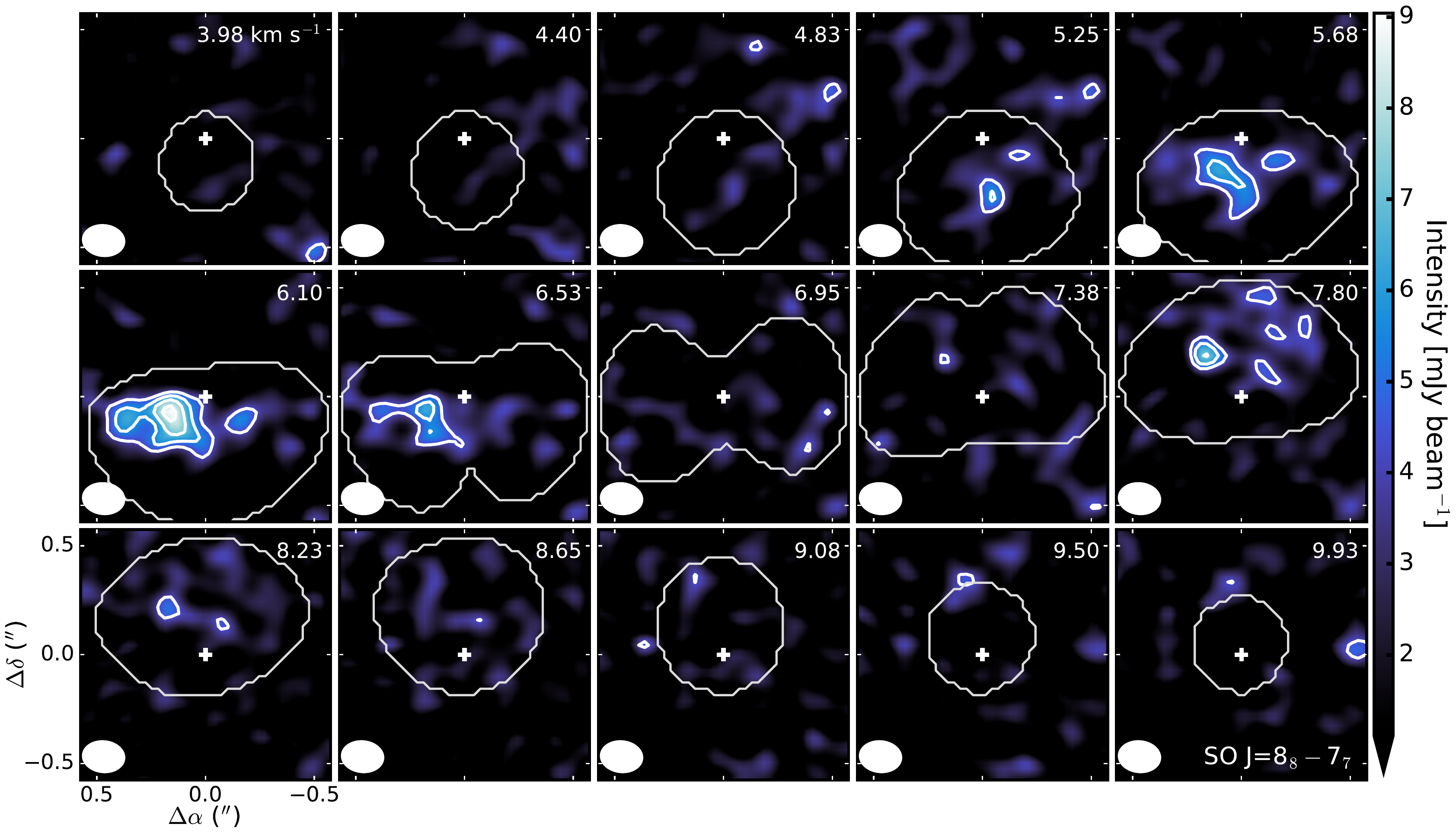}
\caption{Channel maps of the SO J=8$_8-$7$_7$ emission of the HD~169142 disk. A representative Keplerian mask is shown in light gray. Contours show RMS$\times$[3,4,5,6]. The synthesized beam is shown in the lower left corner of each panel and the LSRK velocity in km~s$^{-1}$ is printed in the upper right.}
\label{fig:figure_appendix_so}
\end{figure*}

\begin{figure*}[p!]
\centering
\includegraphics[width=\linewidth]{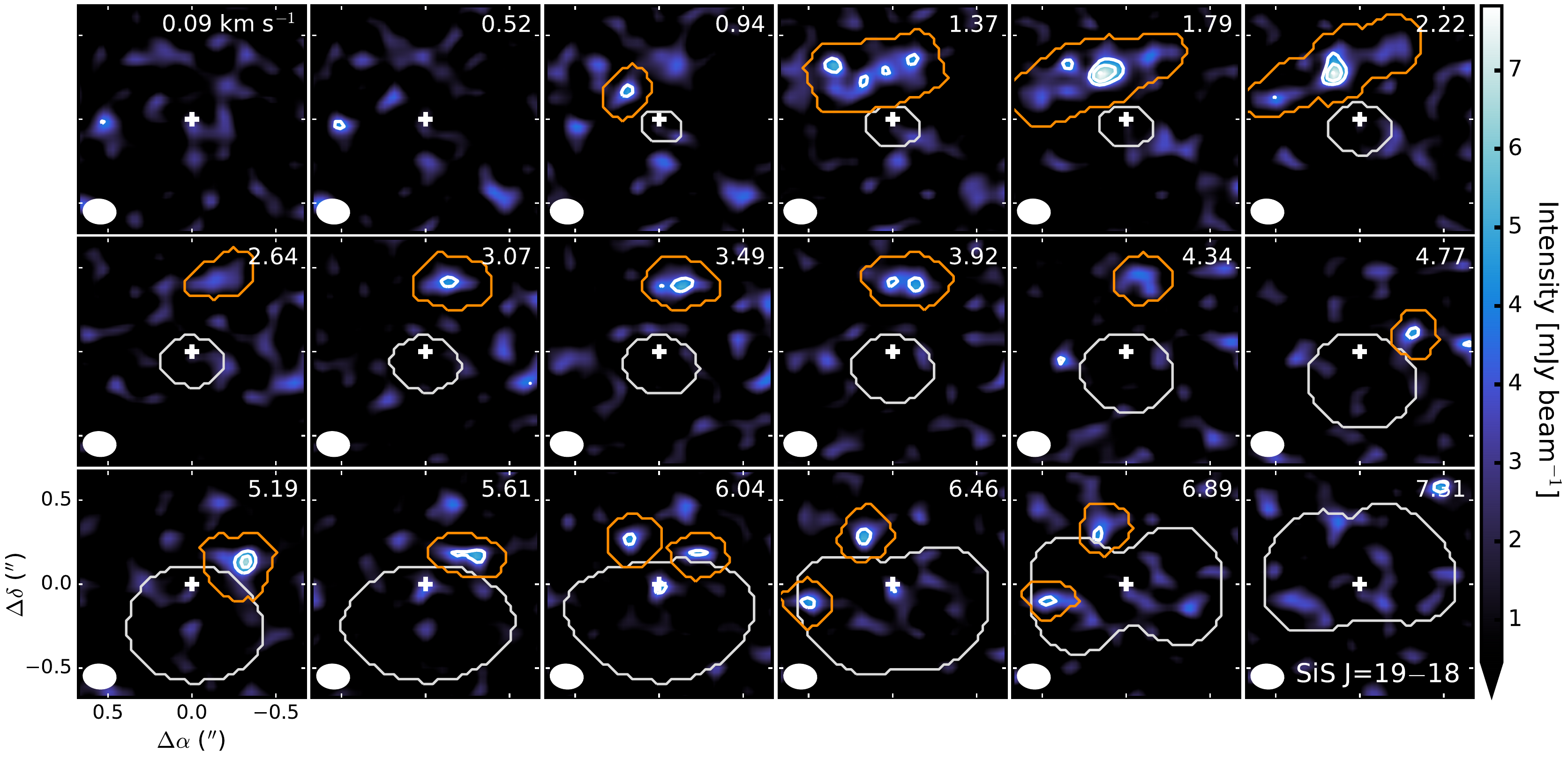}
\caption{Channel maps of the SiS J=19$-$18 emission of the HD~169142 disk. Contours show RMS$\times$[3,4,5]. The hand-drawn masks are marked in orange. Otherwise, as in Figure \ref{fig:figure_appendix_so}.}
\label{fig:figure_appendix_sis}
\end{figure*}

Channel maps of SO J=8$_{8}$--7$_{7}$ and SiS J=19--18 are shown in Figures \ref{fig:figure_appendix_so} and \ref{fig:figure_appendix_sis}, respectively.

\section{Centroid Measurements of Compact $^{12}$CO and $^{13}$CO Emission} \label{sec:CPD_Centroids}

We computed the centroid of the point-source-like emission in $^{12}$CO and $^{13}$CO J=3--2 on a per-channel basis. We used the \texttt{centroid\_2dg} task in \texttt{Photutils} \citep{larry_bradley_2022_6825092} python package to fit a 2D Gaussian to the emission distribution. Table \ref{tab:Centroid} reports the derived centroids.

\begin{deluxetable*}{lccccc}
\tablecaption{Centroid Positions of Compact Emission\label{tab:Centroid}}
\tablewidth{0pt}
\tablehead{
\colhead{Velocity (km~s$^{-1}$)} & \colhead{R.A. (J2000)} & \colhead{decl. (J2000)} & \colhead{R.A. Offset ($^{\prime \prime}$)} & \colhead{decl. Offset ($^{\prime \prime}$)}
}
\startdata
\textbf{$^{12}$CO J=3--2} & \\
3.14  &  18h24m29.7972s & $-$29d46m49.6070s & 0.303 & 0.319 \\
3.26  &  18h24m29.7948s & $-$29d46m49.6043s & 0.272 & 0.322 \\
3.38  &  18h24m29.7947s & $-$29d46m49.6343s & 0.271 & 0.292 \\
3.50  &  18h24m29.7966s & $-$29d46m49.6590s & 0.296 & 0.267 \\
3.62  &  18h24m29.8000s & $-$29d46m49.6642s & 0.340 & 0.262 \\ \hline
\textbf{$^{13}$CO J=3--2} & \\
4.70  &  18h24m29.7936s & $-$29d46m49.5979s & 0.256 & 0.328 \\
4.82  &  18h24m29.7940s & $-$29d46m49.6190s & 0.261 & 0.307 \\
4.94  &  18h24m29.7941s & $-$29d46m49.6362s & 0.262 & 0.290 \\
5.06  &  18h24m29.7929s & $-$29d46m49.6379s & 0.248 & 0.288 \\
5.18  &  18h24m29.7908s & $-$29d46m49.6477s & 0.220 & 0.278 \\
5.30  &  18h24m29.7909s & $-$29d46m49.6451s & 0.221 & 0.281 \\
5.42  &  18h24m29.7906s & $-$29d46m49.6398s & 0.217 & 0.286 \\
\enddata
\tablecomments{Offset positions are computed from the phase center of the observations ($\alpha$=18h24m29.774s, $\delta$=$-$29d46m49.926s).}
\end{deluxetable*}

\begin{figure*}[h!]
\centering
\includegraphics[width=\linewidth]{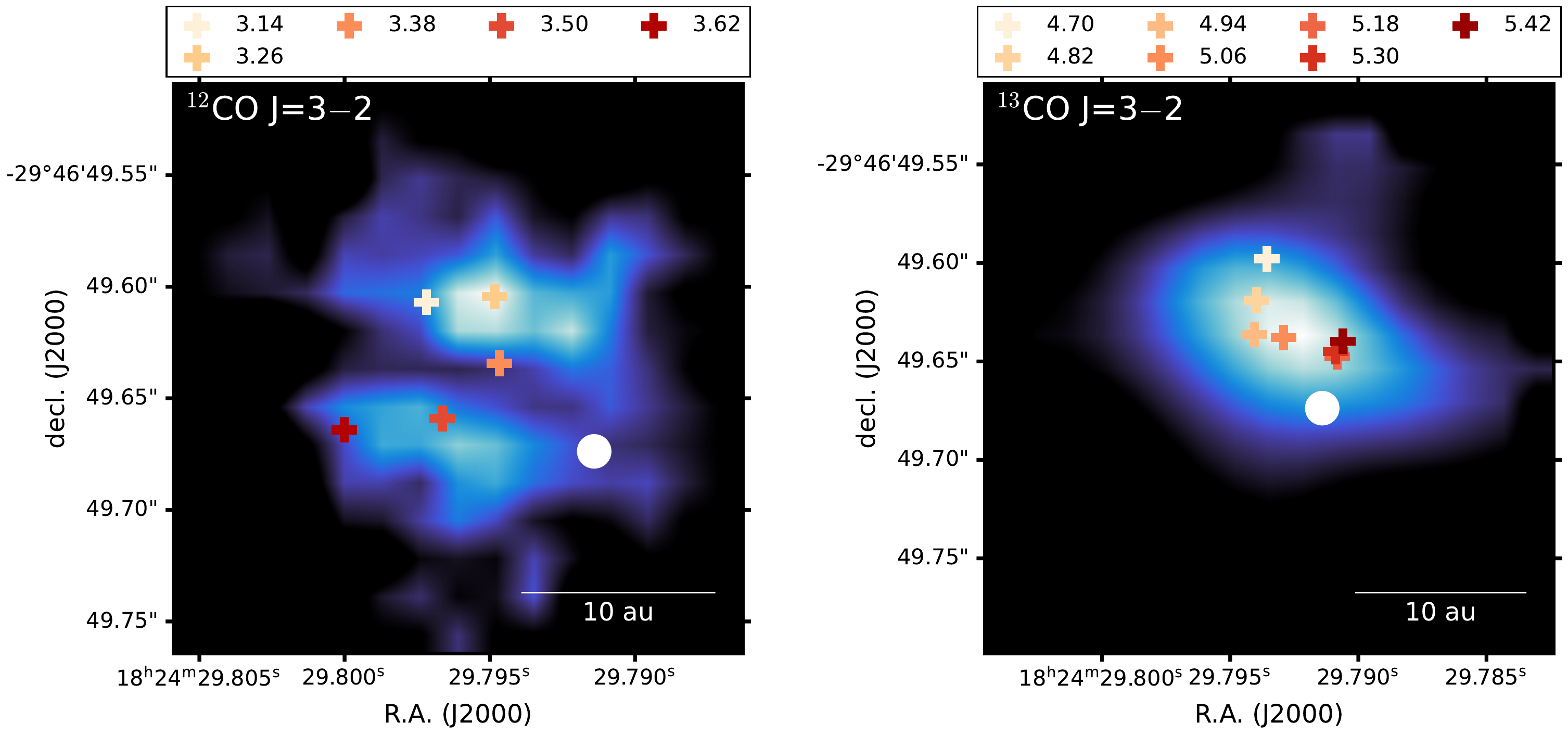}
\caption{Centroid measurements for the compact $^{12}$CO and $^{13}$CO emission over the velocity range in which it is detected. Colors indicate the LSRK velocity in km~s$^{-1}$ at which the centroid was derived. The location of the putative planet is shown by a white circle \citep{Gratton19, Hammond23}. A scale bar showing 10~au is in the lower right.}
\label{fig:centroids}
\end{figure*}

\section{SiS Emission versus CO kinematic excess} \label{sec:SiS_vs_CO_kinematcs}

\begin{figure*}[h!]
\centering
\includegraphics[width=\linewidth]{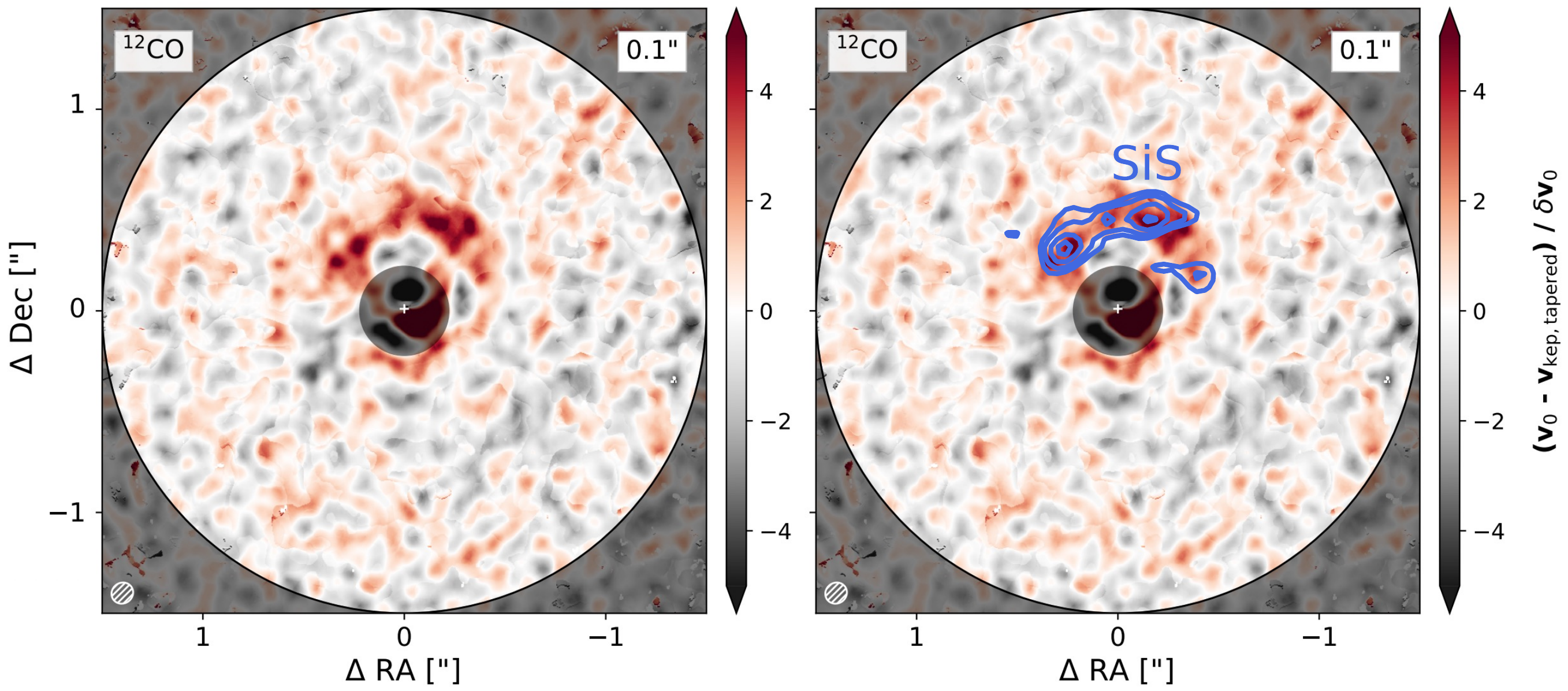}
\caption{$^{12}$CO J=2--1 kinematic excess (left) and with the zeroth moment map of SiS overlaid in contours (right). Background figure reproduced from \citet{Garg22}.}
\label{fig:SiS_vs_CO_kinematcs}
\end{figure*}

Figure \ref{fig:SiS_vs_CO_kinematcs} shows the $^{12}$CO J=2--1 kinematic excess reported in \citet{Garg22} with the integrated intensity of SiS J=19--18 overlaid. The SiS emission is co-spatial with the azimuthal arc of the $^{12}$CO kinematic excess.

\clearpage


\bibliography{HD169142_SO}{}
\bibliographystyle{aasjournal}



\end{document}